\documentclass[aps,prl,10pt,twocolumn,superscriptaddress]{revtex4-2}
\usepackage{graphicx}
\usepackage{amsmath,amsfonts}
\usepackage{color}
\usepackage{ulem}
\newcommand{\upsub}[1]{\sb{\mathrm{#1}}}
\newcommand{\upsup}[1]{\sp{\mathrm{#1}}}

\begingroup\lccode`~=`_\lowercase{\endgroup\let~\upsub}
\begingroup\lccode`~=`^\lowercase{\endgroup\let~\upsup}

\AtBeginDocument{%
	\catcode`_=12 \catcode`^=12
	\mathcode`_="8000
	\mathcode`^="8000
}

\begin{document}


	\title{Bright single photon emitters with enhanced quantum efficiency in a two-dimensional semiconductor coupled with dielectric nano-antennas}

	\author{Luca Sortino}
	\email{luca.sortino@physik.uni-muenchen.de}
	\affiliation{Department of Physics and Astronomy, University of Sheffield, Sheffield, S3 7RH, United Kingdom}
	\affiliation{Chair in Hybrid Nanosystems, Nanoinstitute M{\"u}nich, Faculty of Physics, Ludwig-Maximilians-Universit{\"a}t M{\"u}nchen, 80539 M{\"u}nich, Germany}
	\author{Panaiot G. Zotev}
	\author{Catherine L. Phillips}
	\author{Alistair J. Brash}
	\affiliation{Department of Physics and Astronomy, University of Sheffield, Sheffield, S3 7RH, United Kingdom}
	\author{Javier Cambiasso}
	\affiliation{The Blackett Laboratory, Department of Physics, Imperial College London, London, SW7 2BW, United Kingdom}
	\author{Elena Marensi}
	\affiliation{IST Austria, Am Campus 1, 3400 Klosterneuburg, Austria}
	\author{A. Mark Fox}
	\affiliation{Department of Physics and Astronomy, University of Sheffield, Sheffield, S3 7RH, United Kingdom}
	\author{Stefan A. Maier}
	\affiliation{Chair in Hybrid Nanosystems, Nanoinstitute M{\"u}nich, Faculty of Physics, Ludwig-Maximilians-Universit{\"a}t M{\"u}nchen, 80539 M{\"u}nich, Germany}
	\affiliation{The Blackett Laboratory, Department of Physics, Imperial College London, London, SW7 2BW, United Kingdom}
	\author{Riccardo Sapienza}
	\affiliation{The Blackett Laboratory, Department of Physics, Imperial College London, London, SW7 2BW, United Kingdom}
	\author{Alexander I. Tartakovskii}
	\email{a.tartakovskii@sheffield.ac.uk}
	\affiliation{Department of Physics and Astronomy, University of Sheffield, Sheffield, S3 7RH, United Kingdom}
	
	\date{\today}
	\maketitle

\title{Bright single photon emitters with enhanced quantum efficiency in a two-dimensional semiconductor coupled with dielectric nano-antennas}

\author{Luca Sortino}
\email{luca.sortino@physik.uni-muenchen.de}
\affiliation{Department of Physics and Astronomy, University of Sheffield, Sheffield, S3 7RH, United Kingdom}
\affiliation{Chair in Hybrid Nanosystems, Nanoinstitute M{\"u}nich, Faculty of Physics, Ludwig-Maximilians-Universit{\"a}t M{\"u}nchen, 80539 M{\"u}nich, Germany}
\author{Panaiot G. Zotev}
\author{Catherine L. Phillips}
\author{Alistair J. Brash}
\affiliation{Department of Physics and Astronomy, University of Sheffield, Sheffield, S3 7RH, United Kingdom}
\author{Javier Cambiasso}
\affiliation{The Blackett Laboratory, Department of Physics, Imperial College London, London, SW7 2BW, United Kingdom}
\author{Elena Marensi}
\affiliation{IST Austria, Am Campus 1, 3400 Klosterneuburg, Austria}
\author{A. Mark Fox}
\affiliation{Department of Physics and Astronomy, University of Sheffield, Sheffield, S3 7RH, United Kingdom}
\author{Stefan A. Maier}
\affiliation{Chair in Hybrid Nanosystems, Nanoinstitute M{\"u}nich, Faculty of Physics, Ludwig-Maximilians-Universit{\"a}t M{\"u}nchen, 80539 M{\"u}nich, Germany}
\affiliation{The Blackett Laboratory, Department of Physics, Imperial College London, London, SW7 2BW, United Kingdom}
\author{Riccardo Sapienza}
\affiliation{The Blackett Laboratory, Department of Physics, Imperial College London, London, SW7 2BW, United Kingdom}
\author{Alexander I. Tartakovskii}
\email{a.tartakovskii@sheffield.ac.uk}
\affiliation{Department of Physics and Astronomy, University of Sheffield, Sheffield, S3 7RH, United Kingdom}

\date{\today}
\maketitle

\noindent
\textbf{Single photon emitters in atomically-thin semiconductors can be deterministically positioned using strain induced by underlying nano-structures. Here, we couple monolayer WSe$_2$ to high-refractive-index gallium phosphide dielectric nano-antennas providing both optical enhancement and monolayer deformation. For single photon emitters formed on such nano-antennas, we find very low (femto-Joule) saturation pulse energies and up to 10$^4$ times brighter photoluminescence than in WSe$_2$ placed on low-refractive-index SiO$_2$ pillars. We show that the key to these observations is the increase on average by a factor of 5 in the quantum efficiency of the emitters coupled to the nano-antennas. This further allowed us to gain new insights into their photoluminescence dynamics, revealing the roles of the dark exciton reservoir and Auger processes. We also find that the coherence time of such emitters is limited by intrinsic dephasing processes. Our work establishes dielectric nano-antennas as a platform for high-efficiency quantum light generation in monolayer semiconductors.}

\smallskip
\noindent
\textbf{Introduction}\\
Single photon emitters (SPEs) in two-dimensional (2D) semiconducting WSe$ _2 $ \cite{Tonndorf2015a,Srivastava2015,Koperski2015a,He2015a,Chakraborty2015} open attractive perspectives for few-atom-thick devices for quantum technologies owing to favourable excitonic properties \cite{Wang2017} and the integration with arbitrary substrates, including nano-structured surfaces \cite{Sortino2019,Sortino2020}. Several theoretical models have been proposed to provide insight into the origin of SPEs observed in the cryogenic photoluminescence (PL) spectra of 2D WSe$ _2 $ \cite{Feierabend2019,Lindlau2018,Linhart2019a,Dass2019}. Their occurrence was explained by the presence of strain-induced potential traps for excitons \cite{Feierabend2019}, momentum-dark states \cite{Lindlau2018} or various types of defects \cite{Linhart2019a,Dass2019}. While the exact origin is still under debate, first significant steps have been made to integrate WSe$_{2}$ SPEs in devices, including electroluminescent structures \cite{Palacios-Berraquero2016a,Schwarz2016a,Clark2016}, waveguides \cite{Blauth2018,Peyskens2019} and tunable high-Q microcavities \cite{Flatten2018}.

An appealing approach for the scalable and controllable fabrication of SPEs in WSe$ _2$ is the use of strain engineering for their deterministic positioning. Based on this idea, SiO$ _2 $ \cite{Palacios-Berraquero2016c} or polymer nano-pillars \cite{Branny2016} have been employed to induce arrays of SPEs in atomically thin WSe$ _2 $. In a similar approach,  nano-structures made of noble metals were also employed where, due to the enhancement of the near-field intensity by plasmonic resonances, increased spontaneous emission rates were demonstrated \cite{Luo2018}. However, plasmonic nano-antennas are known for large non-radiative losses, particularly detrimental for quantum technology applications. Thus, special care needs to be taken to separate SPEs from metallic surfaces with a dielectric spacer, which on the other hand will reduce the desired near-field coupling \cite{Luo2018}.

In contrast, the high-refractive-index dielectric materials used in our work offer a lossless alternative to metals \cite{Caldarola2015}. Sub-wavelength dielectric nano-antennas exhibit optical Mie resonances  carrying both electric and magnetic responses \cite{Koshelev2020a}. High-index dielectric nano-antennas have also been recently shown to provide an efficient approach for the enhancement of light-matter interaction as well as improved emitted light directionality in molecules \cite{Cambiasso2017a}, colloidal quantum dots \cite{Kolchin2015} and excitons in 2D semiconductors \cite{Sortino2019,Cihan2018}.

Here, we realise SPEs by placing monolayer WSe$_2$ on top of dielectric nano-antennas made from high-refractive index GaP. Such SPEs show considerably enhanced PL counts per unit excitation power compared with previously reported emitters in WSe$_2$ as well as the SPEs realised on low-index SiO$_2$ nano-pillars in our work. The nano-antennas act as broadband optical cavities and also create strain pockets where the SPEs form. For such SPEs, our numerical simulations predict PL enhancement factors $ \langle EF \rangle $  \cite{Sortino2019,Koenderink2017} up to 800, compared with a more standard realisation of WSe$_2$ SPEs on SiO$ _2 $ pillars \cite{Palacios-Berraquero2016c}. However, owing to the substantially enhanced quantum efficiency ($ QE $) in the SPEs on GaP nano-antennas, we can employ low excitation powers below fJ per laser pulse for their efficient operation. Thus, when we compare SPEs on GaP nano-antennas and SiO$_2$ pillars  experimentally, we find up to 10$^4$ brighter PL per unit laser power in the former system. We show that this substantial improvement in operation of the SPEs is related to the low $ QE $ of the SPEs on SiO$_2$, 4 $\pm$ 2$\%$ on average, compared with 21 $\pm$ 3$\%$ average and up to 86$\%$ maximum in SPEs on GaP nano-antennas. For the latter we find that for the pumping laser repetition rate of 80 MHz, the SPE generates an effective single photons rate as high as 69 MHz under laser pulse energies around a fJ, corresponding to single photon rate of 5.5 MHz at the first lens.

Our approach allows further insight in the exciton dynamics in the hybrid 2D/0D system (2D monolayer/SPE) at very low excitation densities. We observe that as the pumping power is increased, exciton-exciton annihilation \cite{Mouri2014,Danovich2016a}, in our case of dark excitons, prevents efficient population of SPEs. This  insight allows to develop a fuller understanding of the limitations of the low $ QE $ systems, in our case the SPEs on SiO$_2$. There, the requirement for increased pumping power leads to a fast non-radiative depletion of excitons in 2D WSe$_2$ and eventually results in a low single photon generation rate, that cannot be overcome by further increasing the pumping power. Our results thus highlight that the high-refractive-index nano-antennas, exhibiting near-field optical enhancement, provide strong advantages for producing bright SPEs in WSe$_2$ monolayers.

\begin{figure*}[t!]
	\centering
	\includegraphics[width=1\linewidth]{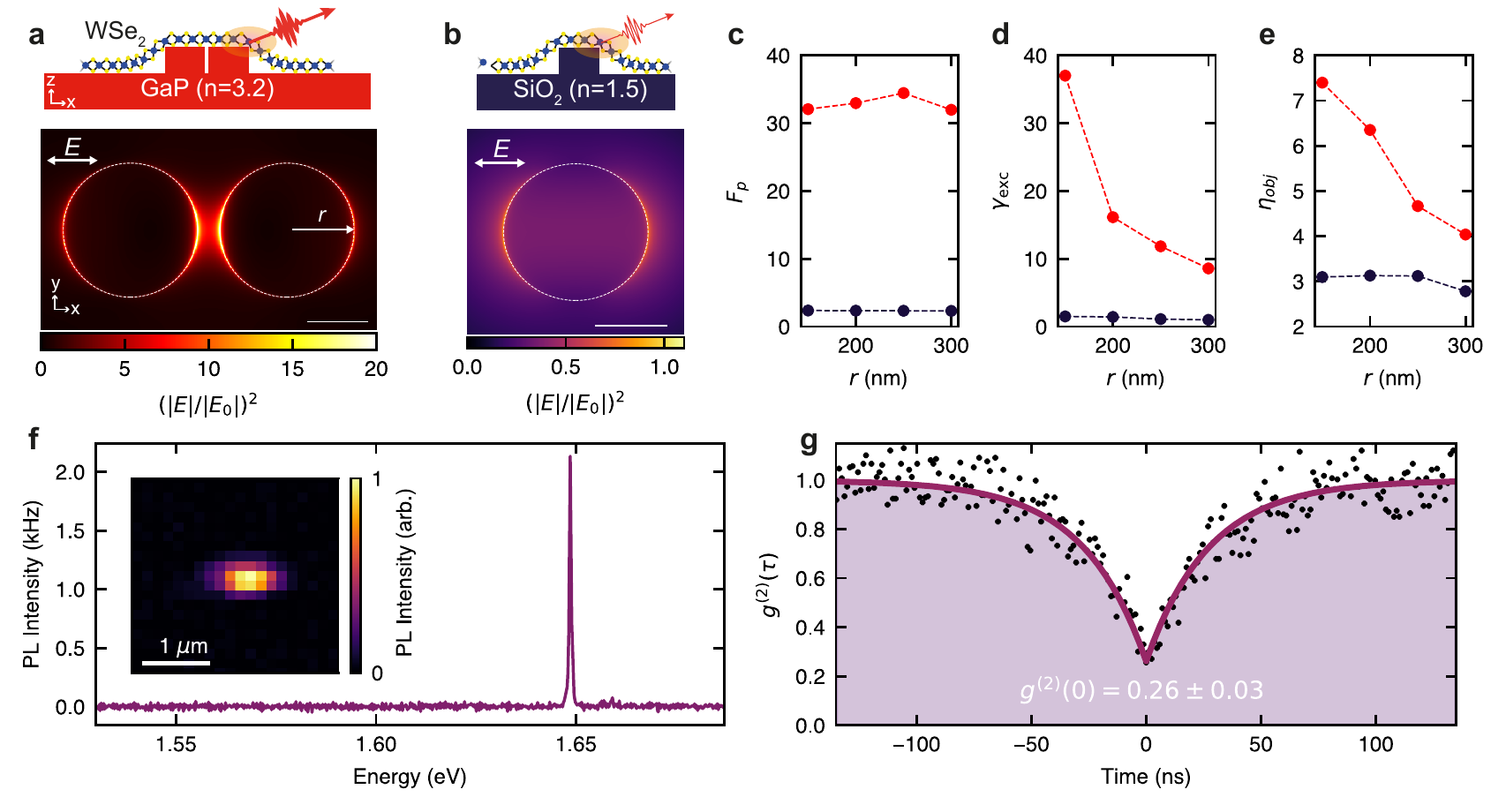}
	\linespread{1}
	\caption{ {\bf Optical properties of nano-antennas and single-photon emitters in monolayer WSe$_2$.}
		(a, b) Top panel: schematics of monolayer WSe$ _2 $ on top of GaP nano-antennas (a) and SiO$_2$ nano-pillar (b). Lower panel: calculated relative intensity of the electric field in the scattered wave ($ E $) over that in the incident wave ($ E_0 $). Results are shown for (a) a GaP dimer nano-antenna ($ r =150 $ nm, $ h = 200$ nm, $gap = 50$ nm) and (b) SiO$ _2$ nano-pillar ($ r = 150 $ nm, $ h = 100$ nm). The calculated intensity is shown for the plane at the top of each structure. Scale bar: 150 nm.
		(c-e) The calculated Purcell enhancement factor (c), excitation rate (d) and light collection efficiency (e) for a dipole emitter placed at the position of the highest field enhancement shown in (a) and (b) of the GaP dimer nano-antennas (red) and SiO$ _2 $ nano-pillars (dark blue) relative to the same parameters for this dipole in vacuum. See further details in Methods and Supplementary Note I.
		(f) A $T= 4$ K PL spectrum of a WSe$ _2 $ SPE on a GaP nano-antenna ($ r=250 $ nm, $ h = 200$ nm, $gap = 50$ nm), excited with a 638 nm pulsed laser with 20 MHz repetition rate and an average power of 15 nW. Inset: Map of the integrated PL intensity of this monolayer ($T= 4$ K). 
		(g) Second order photon correlation curve for the PL signal in (f). }
	\label{fig:fig1}
\end{figure*}

\bigskip
\noindent
\textbf{Results} \\
\textbf{Optical properties of WSe$ _2 $ SPEs positioned on GaP nano-antennas.}
We use GaP nano-antennas composed of two closely spaced nano-pillars (Fig.\ref{fig:fig1}a), referred to as \textquoteleft dimer' below (see also Supplementary Fig.1a). They exhibit an enhancement of the electromagnetic near-field intensity, as a result of the high refractive index ($ n_{GaP}\approx3.2 $) and the hybridization of the optical resonances of each individual pillars (see Supplementary Notes I \& II). This is demonstrated in Fig.\ref{fig:fig1}a, where we show the calculated electric field intensity of the scattered radiation, $ |E|^{2} $, normalized by the intensity $ |E_0|^{2} $ of the normally incident plane wave with linear polarization along the axis connecting the centres of the nano-pillars. The profile in Fig.\ref{fig:fig1}a corresponds to the top surface of the GaP dimer ($ z=200 $ nm) having individual pillar radii of 150 nm. The enhancement of $ |E|^{2} $ compared with $ |E_0|^{2} $ exceeds 10 times and is particularly pronounced in the gap between the pillars \cite{Albella2013,Cambiasso2017a,Sortino2019}. The field is also strongly enhanced at the outer edges of the dimer where we expect SPEs to be located, as discussed further below and in Supplementary Note I. As shown in Fig.\ref{fig:fig1}b, under the same excitation conditions, a SiO$_2 $ nano-pillar ($ r = 150$ nm, $ h = 100 $ nm) does not show strong electromagnetic resonances, as a consequence of its low refractive index ($ n_{SiO_2}\approx1.5 $).

A dipole emitter, such as an exciton in an SPE, spectrally and spatially overlapping with the near-field of the antenna, is expected to exhibit an enhanced light emission intensity \cite{Koenderink2017}. This is a result of the product of the three main factors giving rise to the PL enhancement factor $ \langle EF\rangle $\cite{Sortino2019,Koenderink2017}. Depending on the relative position and orientation of its dipole moment, the emitter experiences an increased local density of states and thus an enhanced spontaneous emission rate \cite{Mignuzzi2019} via the Purcell effect, introducing a factor $ F_p $, directly improving the overall quantum efficiency, $ QE = F_p\gamma_{rad}/(F_p\gamma_{rad}+\gamma_{nr})$, where $\gamma_{rad}$ and $\gamma_{nr}$ are the rates of the radiative and non-radiative decay, respectively. In the absence of the Purcell effect $F_p$=1 and we assumed $\gamma_{rad} \ll \gamma_{nr}$. The antenna also modifies the dipole far field emission pattern, leading to an increased light collection efficiency above the antenna ($ \eta_{obj} $) within the given numerical aperture (NA) of the objective lens used in the detection system. Finally, the enhanced absorption of light in the material coupled to the antenna (in our case, a monolayer WSe$_2$), quantified by the excitation rate proportional to the intensity of the local near-field, $\gamma_{exc}\propto(|E|/|E_0|)^{2}$ (Fig.\ref{fig:fig1}a-b), should in principle lead to a more efficient excitation of an SPE. 

As shown in Fig.\ref{fig:fig1}c-e, we carried out numerical simulations (see Methods and Supplementary Note I) to extract the values of these three parameters for a dipole emitting at $ \lambda_{em} = 750 $ nm coupled to either GaP dimer nano-antennas (data in red) or to SiO$ _2 $ nano-pillars (dark blue). The dipole is placed at the edge of the dimers, where the interaction is maximized \cite{Sortino2019}, and is aligned perpendicularly to the edge of the nano-pillar (Supplementary Fig.1). For the dimers with the radii above 200 nm, the field enhancement at the outer edges is comparable to that in the gap (Supplementary Fig.1a). As shown in Fig.\ref{fig:fig1}c-e, a GaP nano-antenna may induce an enhancement of the PL intensity by at least two orders of magnitude \cite{Albella2013,Cambiasso2017a,Sortino2019}, compared to SiO$ _2 $ nano-pillars, as a consequence of the increase in both the spontaneous emission rate $ F_p $ (Fig.\ref{fig:fig1}c), the excitation rate $\gamma_{exc}$ (Fig.\ref{fig:fig1}d), and a relatively modest effect in the collection efficiency $ \eta_{obj} $ (Fig.\ref{fig:fig1}e), as expected from the similar geometries of dimers and single pillars. 

We note, that due to the high refractive index, for GaP most of the emitted light is directed downwards into the substrate, allowing only about 10$\%$ or less to be collected in the first lens (see further details in Supplementary Note 1). The highest single photon rate of 69 MHz, which we quote in this work, corresponds to the total number of photons emitted by the SPE, whereas in the experiments reported below we measure only up to 5.5 MHz generation rate at the first lens.

In order to experimentally examine these effects, we transferred WSe$ _2 $ monolayers on top of an array of GaP nano-antennas (see Methods and Supplementary Note III) and on SiO$ _2 $ nano-pillars as a reference (see Supplementary Note IV). With this approach, we achieve localized strain in the monolayer, introduced by the underlying nano-structure \cite{Sortino2020}, which promotes the occurrence of localized SPEs at cryogenic temperatures. We find that the SPEs are formed on nearly all nano-antennas with the yield above 90$\%$, in agreement with previous reports \cite{Branny2016,Palacios-Berraquero2016c}. For some nano-antennas we find  multiple emitters (see Supplementary Note III).

The samples are placed in a gas exchange cryostat, at a temperature of $T=4$ K, and excited (unless stated otherwise) with a non-resonant pulsed laser at 638 nm with 90 ps pulse width and a variable repetition rate. The laser excites below the GaP bandgap and is absorbed only by the WSe$ _2 $ monolayer. The inset in Fig.\ref{fig:fig1}f shows a map of the integrated PL intensity from a WSe$ _2 $ monolayer deposited on top of a GaP dimer nano-antenna ($r = 250$ nm). The PL signal exhibits a strong localization at the nano-antenna position, with negligible emission from the surrounding area where the unstrained WSe$_2 $ monolayer is positioned. As shown in Fig.\ref{fig:fig1}f, we observe bright and narrow PL lines, with suppressed background PL from the band of localized states in WSe$_2 $ as is usually observed when WSe$_2$ is deposited on SiO$ _2 $ pillars (Supplementary Fig.7a). We demonstrate the single-photon operation of the localized emitters in photon correlation measurements (see Methods). Fig.\ref{fig:fig1}g shows the second order correlation function, $g^{(2)}(\tau)$, for the emitter shown in Fig.\ref{fig:fig1}f, excited at $ \lambda_{exc}=725 $ nm, approximately 35 meV below the WSe$ _2 $ A-exciton resonance. The pronounced anti-bunching behavior at zero time delay exhibits $ g^{(2)}(0)=0.26\pm 0.03 $, confirming the non-classical photon emission statistics. In Supplementary Note V, we further correlate the SPEs emission energy to the strain induced in the 2D layer by the nano-antennas. Stretching of the WSe$ _2 $ monolayer results in a progressively larger red-shift of the SPE emission when deposited on nano-antennas with smaller radii \cite{Sortino2020}. This behavior, analogous to the red-shift of WSe$ _2 $ excitons under tensile deformation \cite{Sortino2020}, confirms the impact of strain on the confinement potential and emission energy of WSe$ _2 $ SPEs. 

The position of the SPE and the orientation of the emitter dipole relative to the nano-antenna are important factors for the PL enhancement \cite{Sortino2019}. We expect that the SPEs will tend to form naturally around the edges of the nano-pillars (Fig.\ref{fig:fig1}a), where both the tensile strain and photonic enhancement is maximized as follows from our previously reported theoretical and experimental results (see Ref.\cite{Sortino2020} and Supplementary Note V). In these outer edge positions, the SPEs still experience strong enhancement of the electric field, which for the nano-pillars with the radii above 200 nm is comparable with the enhancement in the gap between the pillars (see Supplementary Fig.1a).

In our experiments (not reported here), we find that WSe$_2$ SPEs on planar SiO$_2$ have comparable PL intensities and lifetimes to those formed on SiO$_2$ pillars, with the latter having the advantage of controlled positioning. Furthermore, we did not observe SPEs of reliably measurable PL intensity on planar GaP, where we find that overall PL of WSe$_2$ is quenched as seen for example in Fig.\ref{fig:fig1}f. In what follows we focus on GaP dimer nano-antennas, as these provide pronounced and interesting photonic effects, as was shown in our preliminary work \cite{Sortino2019,Sortino2020}. On the other hand, less studied GaP monomers (single nano-pillars) may also be useful for achieving SPE positioning and improved PL in WSe$_2$ monolayers. This may be a subject of another investigation beyond our current work.

\begin{figure*}[t!]
	\centering
	\includegraphics[width=1\linewidth]{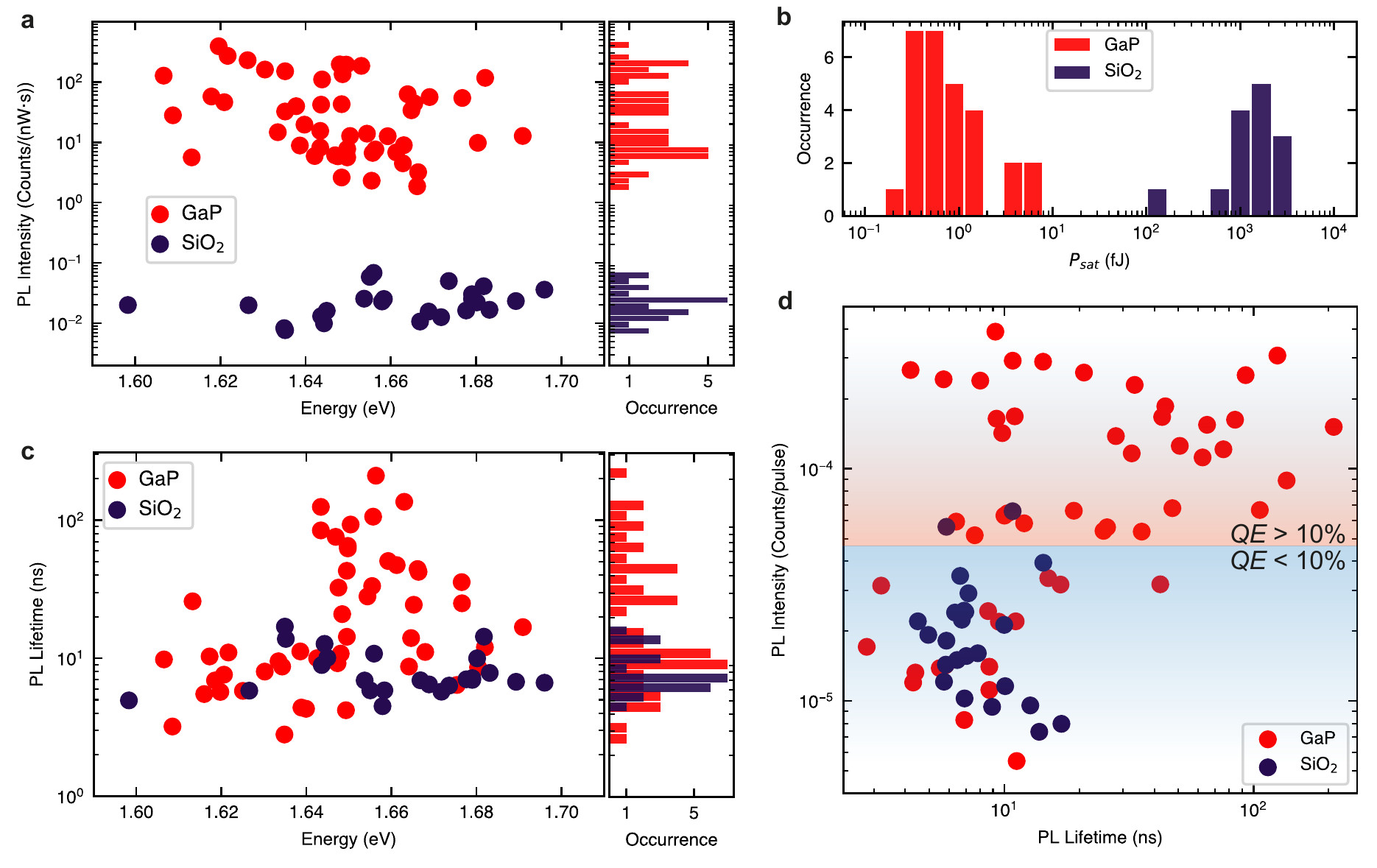}
	\linespread{1}
	\caption{ {\bf Comparison of optical properties of SPEs formed on GaP dimer nano-antennas and on SiO$ _2 $ nano-pillars.}
		Data for SPEs on GaP dimer nano-antennas is shown in red and for SPEs on SiO$ _2 $ nano-pillars is shown in dark blue. 	
		(a) PL intensity normalized by the average laser excitation power measured in the SPE saturation regime. The histogram on the right shows the occurrence of the observed PL intensity values.
		(b) Energy per laser pulse required for SPE saturation, $P_{sat}$. See Supplementary Fig.4 for more details on how $P_{sat}$ is extracted from the PL data.
		(c) PL decay times (main plot) and its occurrence (right).
		(d) SPE PL peak intensity divided by the laser repetition rate plotted versus SPE PL decay time. The red and blue areas of the plot correspond to SPEs with $ QE > 10 \%$ and $ QE < 10 \%$, respectively. See main text and Supplementary Note VI for more details of how $ QE $ was estimated.}
	\label{fig:fig2comparisonsio}
\end{figure*}

\noindent
\textbf{Quantum efficiency enhancement of SPEs on GaP nano-antennas.}
We analysed more than 50 SPEs on GaP dimer nano-antennas, with radii ranging from 150 nm up to 300 nm, selecting localized WSe$ _2 $ emitters with sub-meV linewidths. For these emitters we observed common features such as linearly polarized emission, saturation of the PL intensity under increased excitation power, as well as PL lifetimes in the ns range (see Supplementary Note III). We observed no preferential orientation in the SPEs polarization, with different orientation of polarization even for emitters created on the same nano-antenna.
Fig.\ref{fig:fig2comparisonsio}a shows the values of the PL intensity for SPEs on GaP nano-antennas (red dots) and on SiO$ _2$ nano-pillars (dark blue dots), acquired in the PL saturation regime and normalized to the average excitation pump power. The  plot also shows the PL peak position for each studied SPE, where no correlation between the intensity and spectral position is observed. For SPEs coupled to GaP nano-antennas, we observe from two to four orders of magnitude higher power-normalized PL intensity, compared to SPEs found on SiO$_2$ nano-pillars. Further insight into this behavior is provided by the SPE PL saturation powers, presented in Fig.\ref{fig:fig2comparisonsio}b. Since we used different repetition rates from 5 to 80 MHz in these measurements due to a large variation in the PL lifetimes, in Fig.\ref{fig:fig2comparisonsio}b we plot the energy per pulse value $P_{sat}$, defined as the time-integrated average power divided by the laser repetition rate. We readily observe more than three orders of magnitude lower saturation pulse energies for the emitters on GaP nano-antennas.  In our case, 1 fJ pulse energy corresponds to the energy density per pulse of 30 nJ/cm$^2$. Nonetheless, for such low powers, the SPEs coupled to GaP nano-antennas provide some of the highest counts per second ($>$30,000) so far observed in TMD monolayers. In what follows, we will consider the factors that could contribute to this observation.

One of the obvious factors, expected to contribute to the reduced values of $P_{sat}$ is the enhanced absorption rate ($\gamma_{exc}$) in WSe$ _2 $ monolayers coupled to GaP nano-antennas. However, this can only account for a reduction of the saturation power of up to 40 times as predicted by our simulations (Fig.\ref{fig:fig1}d). A similar maximum enhancement for the power-normalized PL intensity may be expected due to the enhanced $\gamma_{exc}$.  Thus, additional factors need to be considered, mostly linked to the exciton dynamics and $ QE $ of the combined 2D-WSe$_2$/SPE system.  

\begin{figure*}[t!]
	\centering
	\includegraphics[width=1\linewidth]{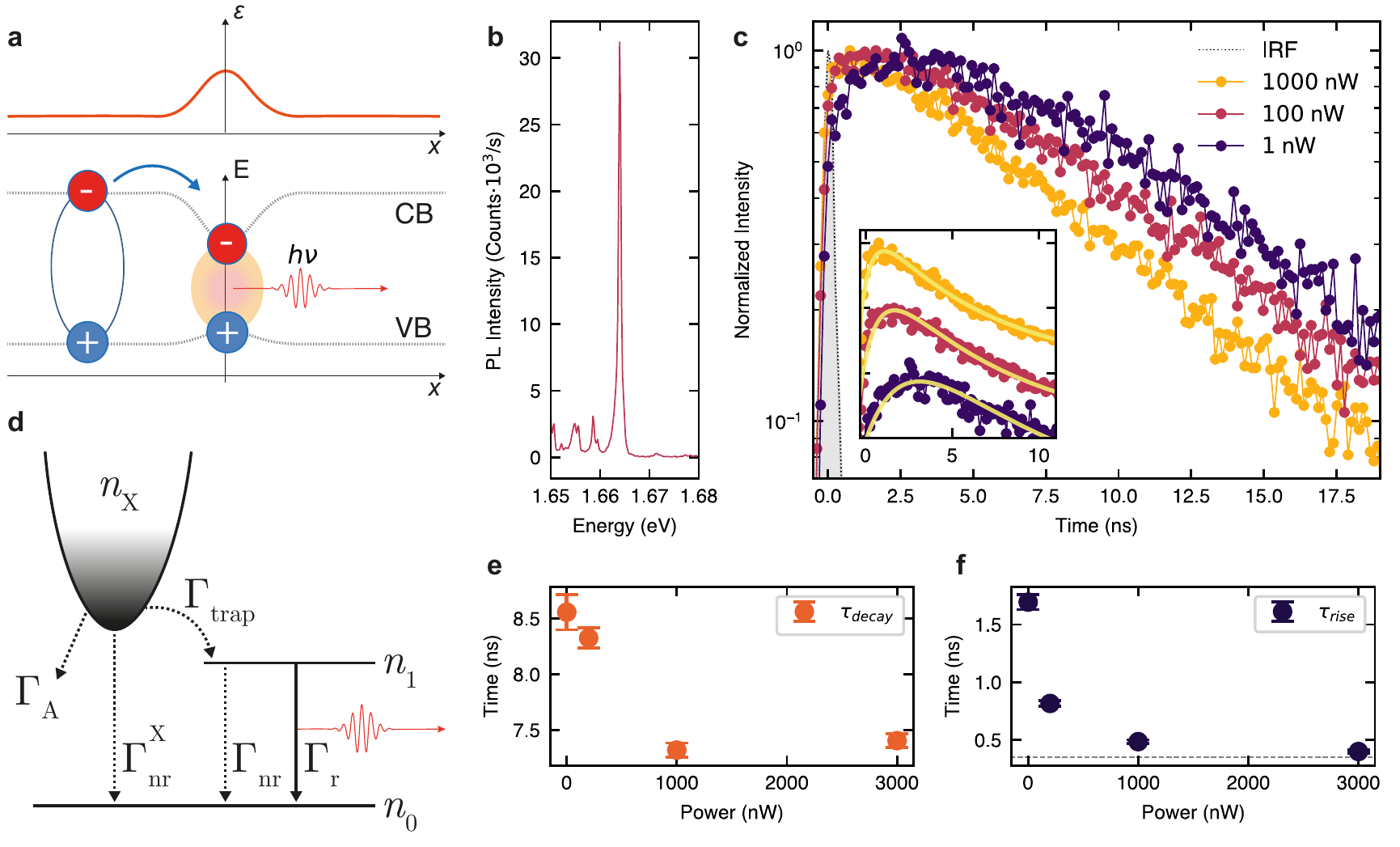}
	\linespread{1}
	\caption{ {\bf PL dynamics in SPEs coupled to GaP nano-antennas.}
		(a) Schematic showing the conduction (CB) and valence band (VB) behavior as a function of strain (shown in the top panel with a red line), and a single exciton trapping from the reservoir in 2D WSe$_2$ into a strain-induced potential minimum, giving rise to the non-classical light red-shifted from the emission in the unstrained monolayer.
		(b) Time-integrated PL spectrum of the strain-induced WSe$ _2 $ SPE with a $ QE $ of 86\% and saturation power of 57 nW for 80 MHz repetition rate, for which the data in (c), (e) and (f) are presented.
		(c) PL decay curves for the SPE in (b) measured at different laser powers. The instrument response function (IRF) is shown in grey. Inset: zoom-in of the PL traces also showing fitting with the analytical model discussed in Supplementary Note VII. The PL traces in the inset are plotted on a linear scale and shifted vertically for display purposes. 
		(d) Schematic of the three level system representing the dark exciton reservoir ($n_X$), the exciton in the SPE ($n_1$) and the ground level ($n_0$), and the processes describing radiative and non-radiative decay of $n_X$ and $n_1$ populations. See text for more details.  
		(e,f) PL decay (e) and PL rise (f) times, as a function of the excitation power, obtained from the data fitting in Fig.3c. For 3000 nW, 
		$\tau_{rise}$ approaches the instrument resolution (grey dashed line). 
		The error bars are calculated form the standard deviation of the fit values.}
	\label{fig:fig3tau}
\end{figure*}

In Fig.\ref{fig:fig2comparisonsio}c we compare PL lifetimes of SPEs on GaP and on SiO$_2$. For the SPEs on the SiO$ _2 $ nano-pillars we observe lifetimes of the order of 10 ns, consistent with previous reports \cite{Palacios-Berraquero2016c,Branny2016}. On the contrary, the SPEs on the GaP nano-antennas exhibit a broad distribution of lifetime values, ranging from 2 ns up to more than 200 ns.
The radiative and non-radiative population decay rates in WSe$_2$ SPEs ($\Gamma_{nr}$ and $\Gamma_{r}$ in Fig.\ref{fig:fig3tau}d) are dependent on the shape and confinement energy of the strain potential, and the PL decay dynamics is defined by the relationship between them: if one of the rates is much higher than the other, it will define the PL decay time.

In order to shed light on the relationship between these rates, in Fig.\ref{fig:fig2comparisonsio}d we plot the SPEs fluorescence lifetime intensity distribution \cite{Morozov2020}. The SPEs on SiO$ _2 $ exhibit low PL emission with relatively short lifetimes (blue area).  In SPEs coupled to GaP nano-antennas we observe either a much higher PL intensity and similar lifetimes, or longer lifetimes with comparable brightness (red area in Fig.\ref{fig:fig2comparisonsio}d). The $ QE $ of an SPE under pulsed excitation can be estimated from the number of detected photons at saturation, divided by the laser repetition rate \cite{Luo2018}. After taking into account the losses of the experimental set-up and the collection efficiency of the nano-antenna from numerical simulations (see additional details in Supplementary Note VI), we estimate an average $ QE $ for SPEs coupled to GaP nano-antennas of $ 21\pm 3\% $, with a maximum value reaching $ 86\%$. For SPEs on SiO$ _2 $ nano-pillars we estimate an average $ QE $ of $ 4\pm 2\% $ consistent with previous reports \cite{Luo2018}.

We thus conclude that the PL decay times of SPEs on SiO$ _2$ nano-pillars are mainly defined by non-radiative processes (i.e. $\Gamma_{nr} \gg \Gamma_{r}$), and that the true radiative lifetimes should by far exceed the measured decay times of $\approx$10 ns. On the other hand, for the SPEs on GaP nano-antennas exhibiting high $ QE $, the lifetimes are mostly defined by the radiative decay (i.e. $\Gamma_{nr} < \Gamma_{r}$ or $\Gamma_{nr} \ll \Gamma_{r}$), which, as we can conclude from Fig.\ref{fig:fig2comparisonsio}d, vary between 2 and 200 ns. Comparing this with the SPEs on SiO$_2$, we can conclude that the high $ QE $ SPEs on GaP exhibiting lifetimes of the order of 10 ns or shorter are most likely affected by the Purcell enhancement increasing their radiative rates, and thus are possibly positioned in the near-field hotspots. 

The high $ QE $ SPEs with PL decay times $>$10 ns clearly must experience much slower non-radiative processes than SPEs on SiO$_2$, as the non-radiative lifetimes must be slower than the measured PL decay times. This is also in contrast to previously reported SPEs coupled to plasmonic structures \cite{Luo2018}, where despite the very large Purcell enhancement and sub-ns PL lifetimes, the maximum $ QE $ of 12.6$\%$ was reported for WSe$_2$ monolayers extracted similarly to our work from bulk crystals grown by chemical vapour transport. This implies high non-radiative rates in this system ($ \Gamma_{nr}>\Gamma_{r} $). On the other hand, SPEs with PL lifetimes in the range of 100 ns were previously observed only in monolayer WSe$_2$ encapsulated in hexagonal boron nitride \cite{Dass2019}, known for suppressing the non-radiative processes. In our case, possible causes for suppression of non-radiative decay could be high surface quality of crystalline GaP structures, or that some SPEs are formed in the suspended parts of the monolayer in proximity to the near-field hotspots and between the pillars \cite{Sortino2020}. We cannot exclude that some of the high $ QE $ SPEs with PL decay times $>$10 ns still experience Purcell enhancement, implying that the true radiative times in some WSe$_2$ SPEs may reach hundreds of ns. Further detailed insight in the PL dynamics is given below in the discussion of Fig.\ref{fig:fig3tau}.

\noindent
\textbf{Dynamics of exciton formation in strain-induced SPEs.}
SPEs in WSe$ _2 $ are attributed to the occurrence of strain-induced local potential minima \cite{Liu2019c,Branny2016}, essentially zero-dimensional (0D), that can host a small number of confined excitons, similar for example to semiconductor quantum dots \cite{Warburton2000}. Contrary to other group-VI TMDs, tensile strain in WSe$_2$ results in the lowering of the conduction band (CB) minimum and the rise of the valence band (VB) maximum, as shown in Fig.\ref{fig:fig3tau}a, both located at the K points in the momentum space \cite{Chang2013a,Sortino2020}. This creates an energy landscape where a very small fraction of the 2D exciton population may be captured into such 0D centres, giving rise to non-classical light emission from confined states, at photon energies lower than that of both bright and dark excitons in unstrained WSe$ _2 $.

As shown in Fig.\ref{fig:fig2comparisonsio}, in the case of WSe$_2$ placed on GaP nano-antennas, both the quantum yield and brightness of the SPEs are greatly enhanced, allowing new insight into the exciton dynamics in this hybrid 2D-0D system. Fig.\ref{fig:fig3tau}b shows a PL spectrum for an SPE exhibiting $ QE $ of $86\pm3\%$. Fig.\ref{fig:fig3tau}c shows the time-resolved PL decay for the same SPE measured with 20 MHz repetition rate. The PL decay curves are obtained at different powers of 1, 100 and 1000 nW considerably below, close and considerably above the saturation power, respectively. For clarity, the inset zooms in on the short times after the laser pulse excitation. At low power we clearly observe a ns-scale rise time, which shortens as the power is increased also accompanied by a relatively weak shortening of the PL decay time. 

We fit the data with a simple empirical model assuming an exciton reservoir, which feeds excitons into the SPE. The model can be solved analytically (see Supplementary Note VII for more details) and is used to fit the data, as shown in the inset of Fig.\ref{fig:fig3tau}c, providing rise and decay times plotted in Fig.\ref{fig:fig3tau}e-f. Here, we see that as the power is increased, the rise time, $\tau_{rise}$, changes strongly from 1.7 ns to times approaching the experimental resolution, whereas the PL decay time $\tau_{decay}$ decreases from 8.5 to 7.4 ns. 

\begin{figure}[t!]
	\centering
	\includegraphics[width=1\linewidth]{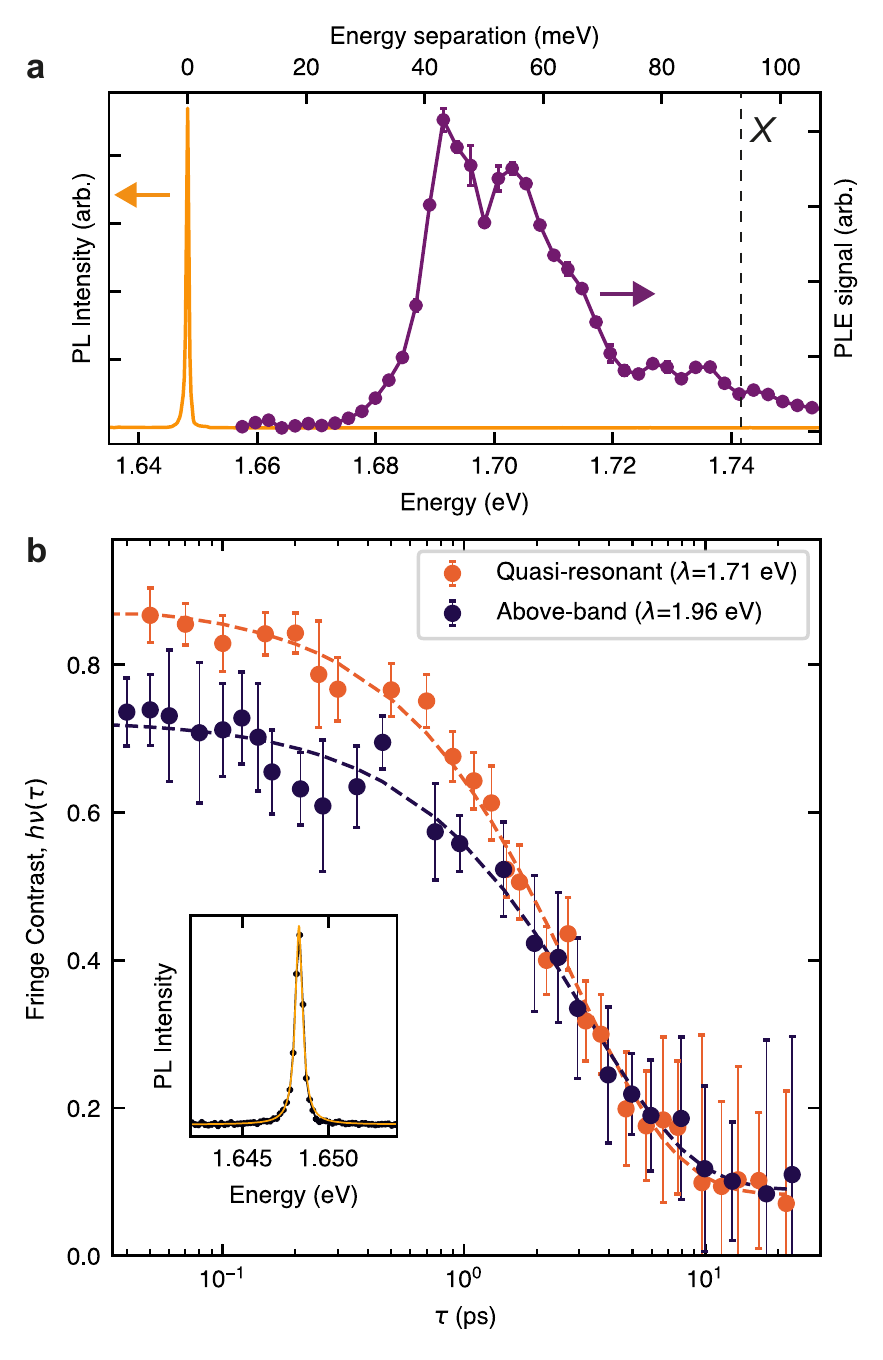}
	\linespread{1}
	\caption{ {\bf Measurements of SPE coherence.}
		(a)  PL emission (yellow) and the PL excitation (PLE) spectra (purple) from a WSe$ _2 $ SPE. The PLE signal exhibits a  series of confined exciton peaks below the neutral exciton PL peak position ($X$).
		(b) Interference fringe contrast, $ \nu(\tau) $ for two different excitation laser wavelengths. The above band excitation (dark blue dots) is obtained at 1.96 eV (638 nm) and yields a coherence time of $ T_2 = 2.83 \pm 0.20 $ (fitting shown with a dashed line). The quasi-resonant excitation (orange dots) is obtained with a laser tuned to 1.71 eV (725 nm), lower than the A-exciton in unstrained monolayer WSe$ _2 $, yielding a coherence time of $ T_2 = 3.12 \pm 0.40 $. The error bars are calculated form the standard deviation of the fit values.
		Inset: SPE PL spectrum under the quasi-resonant excitation (black dots) fitted with a single Lorentzian peak (yellow solid line) with a linewidth of 450 $ \mu $eV, corresponding to a coherence time of $ T_2 = 2.9 $ ps, in close agreement with the values obtained from the interferometric measurements.}
	\label{fig:fig4g1}
\end{figure}

In order to understand this behavior, we consider several processes, which influence both the populations of the high energy 2D exciton reservoir and the SPE itself. We argue that the exciton reservoir with the population $n_X$ in Fig.\ref{fig:fig3tau}d corresponds to the population of dark excitons, which we infer from the very slow PL rise time of 1.7 ns at low power, in contrast to the expected lifetime of the bright excitons of a few ps \cite{Godde2016,Fang2019}. The dark excitons decay mostly via sample-specific non-radiative recombination with a rate $\Gamma^X_{nr}$ and, importantly, via the exciton-exciton (Auger) annihilation \cite{Mouri2014,Danovich2016a}, which grows with the increasing power as $\Gamma_A n^2_X$. Trapping of dark excitons with a rate $\Gamma_{trap}$ into the strain-induced SPE is responsible for a negligible reduction of $n_X$, as the anti-bunching photon emission implies that only one exciton per laser excitation cycle can be created in the SPE. We thus also introduce a probability $n_1$ for the SPE to be filled with an exciton with $0 \leq n_1 \leq 1$. The trapping of the dark excitons is the only source of the SPE population, and is included as a term $\Gamma_{trap}(1-n_1) n_X$ in the equations below. Here we take into account the effect of the SPE occupancy on the reduced efficiency of the exciton trapping with the factor $(1-n_1)$, providing one of the mechanisms for the PL saturation with increasing power observed in the experiment. The population of the SPE decays radiatively and non-radiatively with rates $\Gamma_{r}$ and $\Gamma_{nr}$, respectively. Here, for simplicity we neglect the SPE's internal confined state structure, which we uncover in PL excitation experiments (see Fig.\ref{fig:fig4g1} for details). The rate equations capturing the behavior of the three-level system depicted in Fig.\ref{fig:fig3tau}d are shown below: 

\begin{align}\label{exc}
	\dfrac{dn_X}{dt} &= - [\Gamma^{X}_{nr} +\Gamma_{trap}(1-n_1)]n_X -\Gamma_A n_X^{2}
\end{align}
\begin{align}\label{spe}
	\dfrac{dn_{1}}{dt} &= -(\Gamma_{r} + \Gamma_{nr}) n_{1} + \Gamma_{trap}(1-n_1)n_X
\end{align}

We estimate that for 1 nW laser power at 20 MHz repetition rate and 5$\%$ light absorption in WSe$_2$, the dark exciton density $n_x \approx 3\cdot 10^8$ cm$^{-2}$ will be created. This is probably the lower bound, as the near-field electric field enhancement can locally lead to the increase of this value by a factor exceeding 10. At this low power limit, the Auger annihilation can be neglected \cite{Mouri2014} and the unsaturated SPE emission leads to an average (per pulse) $n_1 \ll 1$. The PL rise dynamics is then defined by the predominately non-radiative decay of the dark exciton reservoir with the rate $\Gamma^X_{nr}$. As the power is increased, and both $n_X$ and $n_1$ grow, two additional processes become important: the Auger annihilation described by the term $\Gamma_A n_X^{2}$ and the saturation of the SPE with the corresponding term $\Gamma_{trap}(1-n_1) n_X$. For the powers presented in Fig.\ref{fig:fig3tau}, $n_X(t=0)$ is estimated to be of the order of $10^{10}$ cm$^{-2}$ for the power of 100 nW and $10^{11}$ cm$^{-2}$ for 1000 nW, in the range where the Auger annihilation was found to be very efficient \cite{Mouri2014,Danovich2016a}. 

While a more detailed study at low powers could help to separate the contributions from the Auger annihilation and SPE saturation, it is possible that in the high power regime the SPE PL saturation is influenced not only by the state-filling effect, but also by the non-radiative depletion of the dark exciton reservoir. In the case of the bright and high $ QE $ SPEs in WSe$_2$/GaP nano-antenna system, high photon counts can be achieved at low excitation powers, thus circumventing the requirement for increased pumping. On the other hand, in the SPEs in WSe$_2$ on SiO$_2$ nano-pillars, where both the $ QE $ and brightness are low, increased pumping is required to observe the SPE PL. This has a negative effect on the population of the reservoir via the Auger annihilation and thus, through such negative feedback, leads to the requirement to further increase the power. Eventually, both the low $ QE $ and its further reduction due to the Auger annihilation lead to a very large three order of magnitude increase in the saturation powers in the SPEs in WSe$_2$ on SiO$_2$ nano-pillars compared with those on GaP nano-antennas, as seen in Fig.\ref{fig:fig2comparisonsio}. In support of these conclusions, we also note that high saturation powers, similar to those observed by us in WSe$_2$ SPEs on SiO$_2$ were also reported in SPEs coupled to plasmonic structures \cite{Luo2018}, where very fast non-radiative processes in the 2D WSe$_2$ should be expected.

\smallskip
\noindent
\textbf{Coherence of a strain-induced SPE.}
The coherence of WSe$ _2 $ SPEs has been previously investigated only under high power densities and non-resonant excitation \cite{Luo2018}. Here, we evaluated the first-order correlation function, $ g^{(1)}(\tau) $, for the SPE shown in Fig.\ref{fig:fig4g1}a, in a Mach-Zender interferometer set-up \cite{Brash2019} and compared different excitation schemes (see Methods). We employed an above-band excitation using a 1.96 eV (638 nm) cw laser, corresponding to an energy higher than the A-exciton resonance in monolayer WSe$ _2$ (dashed line in Fig.\ref{fig:fig4g1}a).  Under these conditions, high energy excitons are created in the continuum of states above the excitonic resonance, introducing dephasing for instance via scattering with phonons and impurities or via exciton-exciton interactions. To reduce the impact of such processes, we also used a quasi-resonant excitation with a cw laser at 1.71 eV (725 nm). As shown in Fig.\ref{fig:fig4g1}a, this excitation is resonant with higher energy states within the SPE \cite{Tonndorf2015a}. 
Fig.\ref{fig:fig4g1}b shows the measured fringe contrast, $ \nu(\tau) $, of the WSe$ _2 $ SPE under the two excitation schemes (see Methods). By fitting the observed decay of the fringe contrast with a single exponential decay function, $ g^{(1)}(\tau) \approx \exp(-|\tau|/T_2) $, we extract a coherence time of $ T_2 = 3.12 \pm 0.40 $ ps under quasi resonant excitation, and of $ T_2 = 2.83 \pm 0.20 $ ps for above band excitation. The differences between the excitation schemes have a negligible effect on the SPE dephasing time, implying a complex relaxation processes within the confined states of the SPE. We find that the PL full-width at half maximum (FWHM) of $ \approx 450$ $\mu$eV corresponds to $ T_2 = 2.9  $ ps (FWHM $= 2\Gamma = 2\hbar / T_2 $) close to the observed $T_2$ values, indicating that the coherence of the studied SPE is limited by pure dephasing, which we attribute to interactions with phonons during the exciton relaxation \cite{Dey2016}, as for the excitation power $ <20 $ nW used in the experiment the contribution of the Auger annihilation can be excluded. The observed SPE $ T_2 $ values are one order of magnitude higher than those reported for monolayer WSe$ _2 $ of 0.3 ps \cite{Dey2016}. Excitation in resonance with the lowest energy optical transition in the SPE could be employed to gain access to the intrinsic coherence times of the confined excitons.

\bigskip
\noindent
\textbf{Discussion}

\noindent
In summary, we have demonstrated that high-refractive-index GaP nano-antennas offer an efficient approach for nano-scale positioning and $ QE $ enhancement in strain-induced SPEs in monolayer WSe$_2$. We found 10$^2$ to $10^4$ enhancement of the PL intensity for WSe$_2$ SPEs coupled to GaP nano-antennas compared with those formed on low-refractive-index SiO$_2$ nano-pillars. We demonstrate that this is primarily caused by the greatly increased $ QE $ in the 2D/0D WSe$_2$ system coupled to GaP nano-antennas arising from the enhancement of the radiative rates in such SPEs through the Purcell effect, as well as the reduction of the non-radiative decay rates. Importantly, this allows bright emission from the SPEs to be excited with energy densities per laser pulse below 30 nJ/cm$^2$ corresponding to the energy per pulse below 1 fJ, enabling the SPE operation at low exciton densities in the 2D WSe$_2$, thus avoiding the exciton-exciton annihilation. The powers at which SPEs on GaP nano-antennas provide bright emission are approximately three orders of magnitude below those required for operation of the SPEs on SiO$_2$ pillars studied in this work, as well as those previously reported for SPEs formed on plasmonic nano-structures \cite{Luo2018}, despite the large Purcell enhancement factors found in the latter system \cite{Luo2018}. Further improvement and consistency of the operation of SPEs can possibly be achieved by employing deterministic defect placement \cite{Klein2020,Parto2021}, while the required excitation powers can be further reduced by employing much cleaner WSe$_2$ grown by the so-called flux technique \cite{Luo2018}. The photon collection efficiency in our approach can be potentially improved by engineering the nanocavity geometry and materials, for instance by using nano-antennas made from high-index TMDs \cite{Zotev2021,Verre2018}. These materials can be deposited on any type of substrate. By placing them on a metallic mirror similarly to the strategy pursued by Luo et al in Ref.\cite{Luo2018}, most of light can be redirected upwards, and the collection efficiency in the first lens can be considerably increased. Overall, our work suggests that hybrid systems composed of 2D semiconductors coupled to dielectric nano-antennas are a powerful means for controlling quantum light generation.

\bigskip
\noindent
\textbf{Methods}

\noindent
\textbf{Sample fabrication}
The GaP dimer nano-antennas were fabricated using electron beam lithography, followed by several wet and dry etching steps as described in Ref.\cite{Cambiasso2017a}. Arrays of nano-antennas separated by 4 $ \mu $m were made. The dimers had a gap of $ \approx50 $ nm, a height of 200 nm and nano-pillar radii ($ r $) of 150, 200, 250 and 300 nm. Atomically thin monolayers of WSe$ _2 $ were mechanically exfoliated from commercially available bulk single crystals (HQ Graphene) onto polydimethylsiloxane (PDMS) polymer substrates. The monolayer thickness was identified by examining room temperature PL with an imaging method described in Ref.\cite{Alexeev2017}. The monolayers were then transferred on top of the GaP nano-antenna array, by using the same PDMS substrates, with an all-dry transfer technique in a home-built transfer setup \cite{Castellanos-Gomez2014a}. 

\smallskip
\noindent
\textbf{Optical spectroscopy}
Low temperature PL spectroscopy was carried out with a sample placed in low pressure He exchange gas within a confocal microscope platform allowing free space optical access and high precision sample positioning (Attocube). The whole microscope stick was inserted in a liquid helium transport dewar (Cryo Anlagenbau Gmbh) and a nominal sample temperature of 4 K was used in all reported experiments. The excitation from the lasers used in this work was delivered through single-mode fibres to the optical breadboard placed at the top of the microscope stick, where it was collimated and directed onto the sample through a window at the top of the stick. For pulsed excitation we used a diode laser (PicoQuant) at 638 nm, with a variable repetition rate from 5 to 80 MHz and a pulse width of 90 ps. For continuous wave excitation, we used a tunable Ti-Sapphire laser (M Squared SOLSTIS). 
PL emitted by the sample was collected  with an aspheric lens (NA = 0.64) and coupled at the breadboard into a single-mode optical fibre, which delivered it to a spectrometer (Princeton Instruments SP2750), where it was detected with a high-sensitivity liquid nitrogen cooled charge-coupled device (Princeton Instruments PyLoN). For the time-resolved spectroscopy, the PL was also sent through the spectrometer to another exit port, where it was measured with an avalanche photodiode (ID100-MMF50) connected to a photon counting card (Becker and Hickl SP-130). 
A Hanbury Brown-Twiss set-up used for the evaluation of the second-order correlation function ($ g^{(2)}(\tau) $) was equipped with two superconducting nanowire single photon detectors (Single Quantum) and a similar photon counting card. The SPE was excited with a continuous wave diode laser (Thorlabs HL7302MG) at 730 nm with a power of 2 nW. Light from the SPE was directed to the nanowire detectors with a multi-mode fibre. The emitted light from the SPE was filtered with a pair of filters (Thorlabs FESH800 and Thorlabs FELH750) providing a transparency window between 750 nm and 800 nm.

\smallskip
\noindent
\textbf{Coherence measurements}
For evaluation of the first-order correlation function ($ g^{(1)}(\tau) $), we used a Mach-Zender interferometer set-up \cite{Brash2019} with a phase shifter in one arm and a variable optical delay in the other. By sweeping the voltage of the phase shifter, the interference of light emitted by the SPE was measured using an avalanche photodiode at one output port of the interferometer. By measuring the intensity at the local maxima ($ I_{max} $) and minima ($ I_{min} $) of the interference fringes, we evaluate the fringe contrast ($ \nu $) as:
\begin{equation}
	\nu = \frac{I_{max}-I_{min}}{I_{max}+I_{min}}
\end{equation}
\noindent
This procedure was repeated for increasing delay times, until the fringes were no longer resolved. The relationship between the fringe contrast and the first-order correlation function is given by the following equation:
\begin{equation}
	\nu(\tau) = (1-\epsilon)\frac{|g^{(1)}(\tau)|}{g^{(1)}(0)}
\end{equation}
\noindent
where $ 1-\epsilon $ is the maximum resolvable fringe contrast in the set-up and $ |g^{(1)}(\tau)| $ is the first order correlation function excluding the fast oscillations at the emitter frequency. \cite{Brash2019}. In the absence of any spectral diffusion, the fringe contrast as a function of  time follows a single exponential decay, with an exponential fit allowing the evaluation of the coherence time ($ T_2 $) of the emitter.

\smallskip
\noindent
\textbf{Simulations}
The distributions of the electric field in Fig.\ref{fig:fig1} were calculated with a commercial finite-difference time-domain software (Lumerical Inc.). In the simulations we illuminated the structure with a linearly polarized plane wave at $ \lambda = $750 nm with a normal incidence from the vacuum side of the substrate. See Supplementary Note I for further details on the simulations.

\bigskip
\noindent
\textbf{Data availability}

\noindent
The data that support the findings of this study are available from the corresponding author upon request.

\smallskip
\noindent
\textbf{Acknowledgments}
L. S., P. G. Z. and and A. I. T. thank the financial support of the European Graphene Flagship Project under grant agreements 881603 and EPSRC grant EP/S030751/1. L. S. and A. I. T. thank the European Union's Horizon 2020 research and innovation programme under ITN Spin-NANO Marie Sklodowska-Curie grant agreement no. 676108. P. G. Z. and A. I. T. thank the European Union's Horizon 2020 research and innovation programme under ITN 4PHOTON Marie Sklodowska-Curie grant agreement no. 721394. J. C., S. A. M., and R. S. acknowledge funding by EPSRC (EP/P033369 and EP/M013812). C. L. P., A. J. B., A. I. T. and A. M. F. acknowledge funding by EPSRC Programme Grant EP/N031776/1. S. A. M. acknowledges the Lee-Lucas Chair in Physics,  the Solar Energies go Hybrid (SolTech) programme, and the Deutsche Forschungsgemeinschaft (DFG, German Research Foundation) under Germany's Excellence Strategy – EXC 2089/1 – 390776260.

\smallskip
\noindent
\textbf{Author contributions}
L. S., P. G. Z., A. I. T., S. A. M. and R. S. conceived the idea of the experiment. L. S. and P. G. Z. fabricated WSe$_2$ layers, transferred them on GaP nano-antennas, and carried out numerical modelling. L. S., P. G. Z., C. L. P. and A. J. B.  carried out optical spectroscopy measurements on WSe$_2$. J. C. fabricated GaP nano-antennas. J. C. and R. S. designed GaP nano-antennas. L. S. and E. M. designed and analysed the rate equation model. L. S., P. G. Z., C. L. P., A. J. B. and A. I. T. analysed optical spectroscopy data. S. A. M., R. S., A. M. F. and A. I. T. managed various aspects of the project. L. S., P. G. Z. and A. I. T. wrote the manuscript with contributions from all co-authors.  A. I. T. oversaw the whole project.

\smallskip
\noindent
\textbf{Competing Interests}
The authors declare no competing interests.

\clearpage

\onecolumngrid
\appendix
\renewcommand{\figurename}{SUPPLEMENTARY FIGURE}
\setcounter{figure}{0}

\section*{Supplementary information for: Bright single-photon emitters with enhanced quantum efficiency in a two-dimensional semiconductor coupled with dielectric nano-antennas}

\section*{Supplementary note I: Numerical simulations of the Photoluminescence Enhancement Factor}

The PL intensity collected from a single dipole emitter coupled to an optically driven nano-antenna is highly dependent on its relative position and orientation in respect to the scattered field, defined by the vector $ \textbf{r} $, and originates from three factors \cite{Koenderink2017}:
\begin{equation}\label{ef}
	I(\textbf{r},\lambda_{em})\propto\gamma_{exc}(\textbf{r},\lambda_{exc})\cdot QE(\textbf{r},\lambda_{em})\cdot\eta_{NA}(\textbf{r}, \lambda_{em})
\end{equation}
The relative PL enhancement factor $ \langle EF \rangle $ \cite{Sortino2019} is given by the ratio between PL intensity values obtained when the dipole is placed on GaP nano-antennas and on SiO$ _2 $ nano-pillars, $ \langle EF \rangle = {I_{GaP}}/{I_{SiO_{2}}} $.
\noindent 
We carried out a set of numerical finite-difference time-domain (FDTD) simulations for an in-plane dipole emitting at $\lambda_{em}$ = 750 nm, placed either on top of GaP dimer nano-antennas, or on SiO$ _2 $ nano-pillars on a SiO$ _2 $(100 nm)/Si substrate. The heights of the GaP dimers and the SiO$_2$ pillars were set to 200 nm and 100 nm, respectively, matching the structures used in our experiments. In our simulations we find that by increasing the height of SiO$_2$ pillars, a negligible effect on their optical properties is observed due to the lack of Mie optical resonances.

\begin{figure}[b!]
	\centering
	\includegraphics[width=1\linewidth]{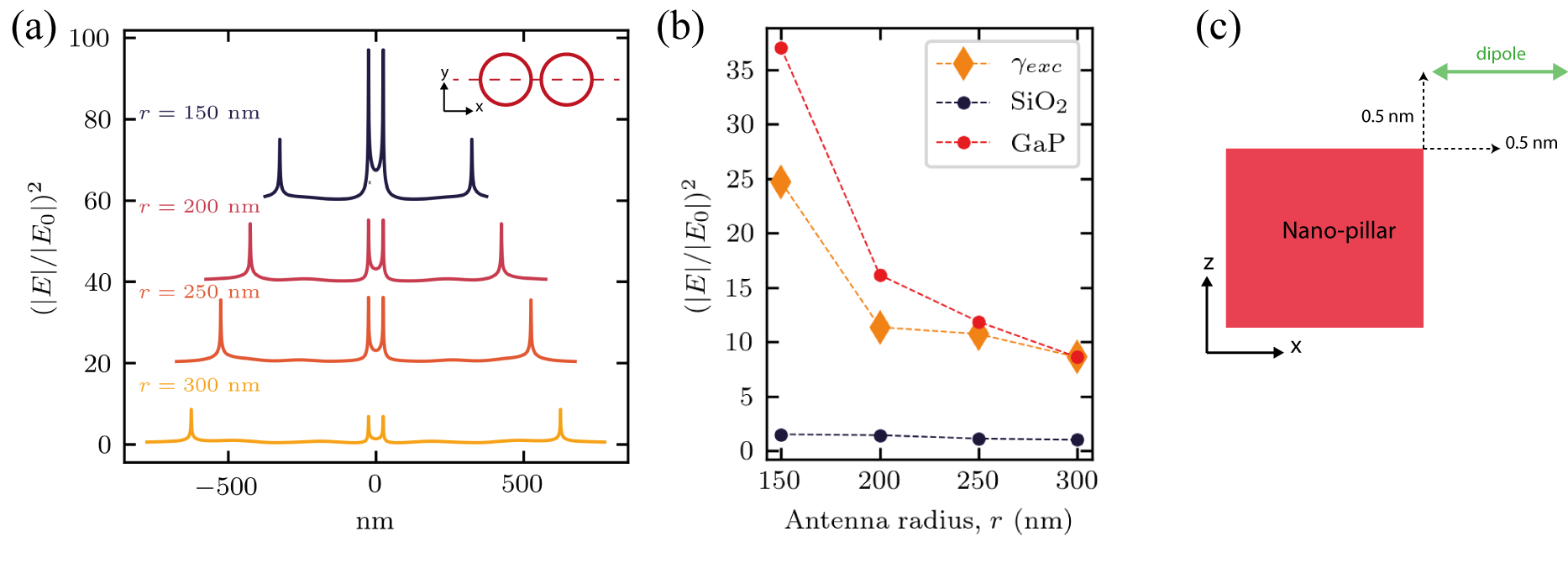}
	\caption{(a) Profiles of the electric field intensity, $ (|E|/|E_0|)^2$, at the top surface of the nano-antenna (z=200 nm) under a plane wave excitation at $ \lambda_{exc} = 638 $ nm linearly polarized along the x-axis. The profile is taken along the axis connecting each nano-pillar centre (dashed line in Inset). The profiles are shifted vertically for display purposes. (b) Values of the maxima of the electric field intensity for GaP dimer nano-antennas (red), SiO$ _2 $ nano-pillars (blue) and their ratio $ \gamma_{exc} = \gamma_{exc}^{GaP}/\gamma_{exc}^{SiO_2}$ (orange). (c) Position of the dipole emitter relative to the nano-pillar edge used in the Purcell factor simulations.}
	\label{fig:s1efexc}
\end{figure}

The first factor in Eq.\ref{ef}, $ \gamma_{exc} $, describes the increased absorption cross section, dependent on the local electric field intensity $ (|E|/|E_0|)^2 $, where $ |E|^{2} $ is the intensity of the scattered radiation and $ |E_0|^{2} $ the intensity of the normally incident linearly polarized plane wave. Supplementary Fig.\ref{fig:s1efexc} shows the maxima of the local E field at the top surface of the nano-antenna ($z = 200$ nm) taken along the line connecting each nano-pillar centre (see Inset). In Supplementary Fig.\ref{fig:s1efexc}b we show the maximum value of the field on GaP dimers (red) and on SiO$ _2 $ nano-pillars (blue). Their ratio (yellow) defines the relative field enhancement, equal to $ \gamma_{exc} $. 

The second factor $ QE $ is the quantum efficiency of the dipole, defined as $ QE = \gamma_{r}/(\gamma_{r} + \gamma_{nr} )$, where $ \gamma_{r} $ is the radiative decay rate and $ \gamma_{nr} $ is the non-radiative decay rate. In our simulations we used an approximation for low $ q $ emitters, where the non-radiative decay $\gamma_{nr} \gg F_p \gamma_r$, and the change in $ q $ can be evaluated only from the radiative decay rate enhancement defined by the Purcell factor \cite{Sortino2019}, $ F_P = \gamma_{r}/\gamma_{r}^{0} $, defined as the enhancement of the energy dissipation $ P/P_0 $ in the numerical simulations \cite{Novotny2006}, where $ \gamma_{r}^{0}  $ and  $ P_0 $ ar related to the dipole on planar substrate.  The dipole is placed 0.5 nm above the surface and 0.5 nm away from the nano-pillar edge, as shown in Supplementary Fig.\ref{fig:patternrad}c. The values obtained are shown in the main text in Fig.1c, where they are normalized over the same dipole placed on the flat substrate as reference.

The last factor in Eq.\ref{ef}, $ \eta_{NA} $, defines the fraction of the light collected by the numerical aperture ($ NA$) of the objective, calculated as the fraction of power emitted in the upwards direction, in a cone defined by the objective $ NA $. Supplementary Fig.\ref{fig:patternrad}a-c shows the radiation pattern for an in-plane dipole ($\lambda_{em}$ = 750 nm) placed at the centre of the gap of a GaP dimer nano-antenna (in red) and at the edge of a SiO$ _2$ nano-pillar (in grey). In case of GaP nano-antennas, no significant difference was observed when placing the dipole at the edge of the nano-pillar, similarly to SiO$ _2 $. Due to the higher refractive index, for GaP most of the emitted light is directed downwards into the substrate. For a dipole coupled to GaP dimer nano-antennas we obtained a collection of $ 7\% $ for antennas with a radius of 300 nm, and up to $ 10\% $ for $ r = 150 $ nm. For a dipole on top of a SiO$ _2 $ nano-pillar, we found a collection from $ 8\%$ (r = 300 nm) up to $9\% $ (r = 150 nm).

In case of dimer nano-antennas larger than 150 nm we observe a larger enhancement of the collection efficiency and $ F_P $ compared to the same dipole placed in the gap of the dimer. Following our previously published theoretical calculations \cite{Sortino2020}, this is the point where tensile strain is maximised and thus where the SPEs will likely be positioned. In Supplementary Fig.\ref{fig:patternrad}d-e we show the effect on the power radiated by a dipole placed at either the gap (purple) or at the outer edges (yellow) of a dimer nano-antenna with a radius of 200 nm.

\begin{figure}[h!]
	\centering
	\includegraphics[width=1\linewidth]{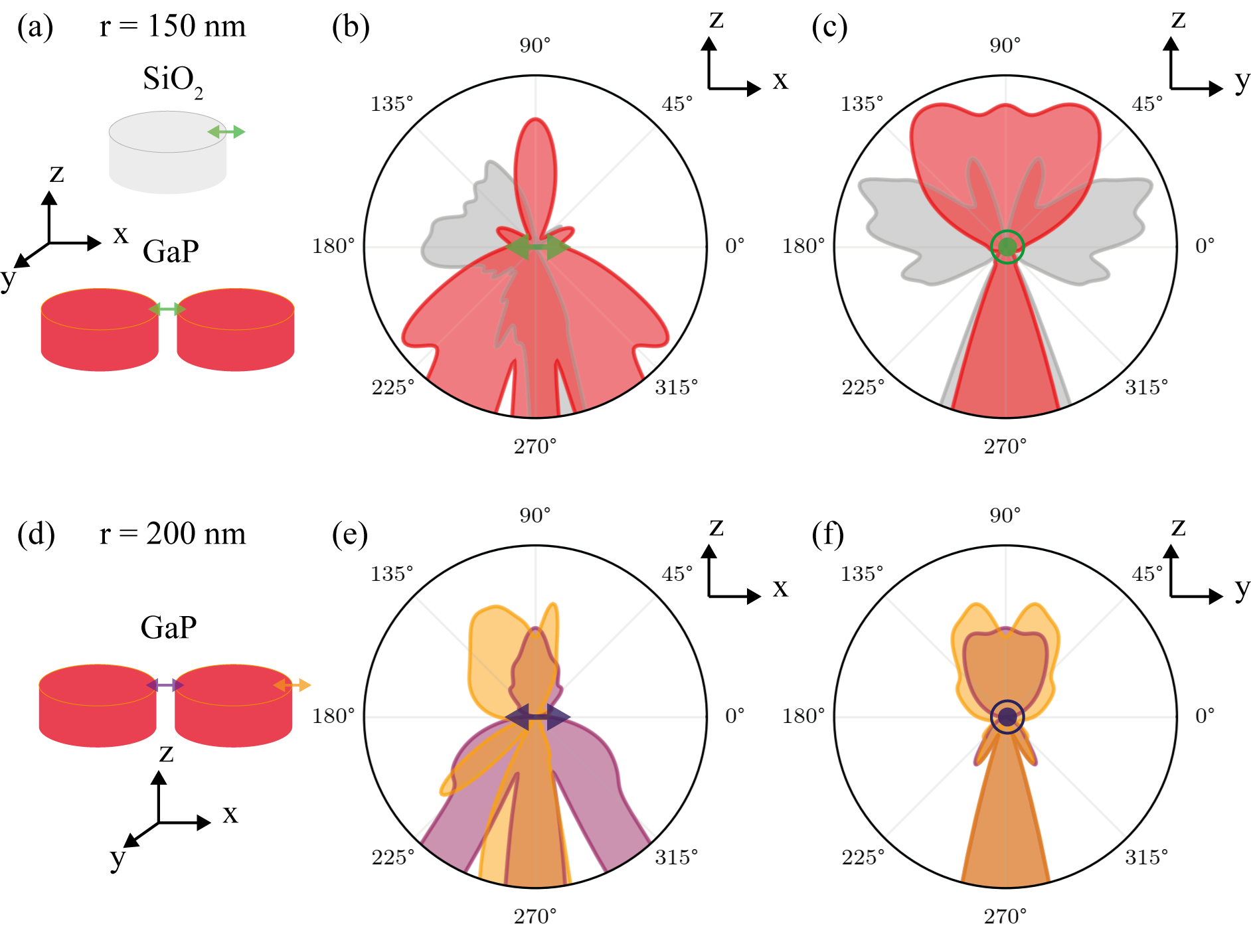}
	\caption{(a) Schematics of the electric dipole (in green) position and orientation in respect to either a silicon nano-pillar (grey) and a GaP dimer nano-antennas, both with a radius of 150 nm. (b,c) Radiation pattern projection on the $ xz $-plane (b) and $ yz $-plane (c) for a dipole closely coupled to a SiO$ _2 $ nano-pillar (in grey) and in the gap of a GaP dimer nano-antenna (in red). (d) Schematics of an electric dipole position and orientation for a GaP dimer nano-antenna with radius of 200 nm. (b,c) Radiation pattern projection on the $ xz $-plane (e) and $ yz $-plane (f) for a dipole placed in the gap of the dimer (in purple) and at the outer edge (in yellow).}
	\label{fig:patternrad}
\end{figure}

\clearpage

\section*{Supplementary note II: Optical Mie resonances of GaP dimer nano-antennas}
\begin{figure}[h!]
	\centering
	\includegraphics[width=0.8\linewidth]{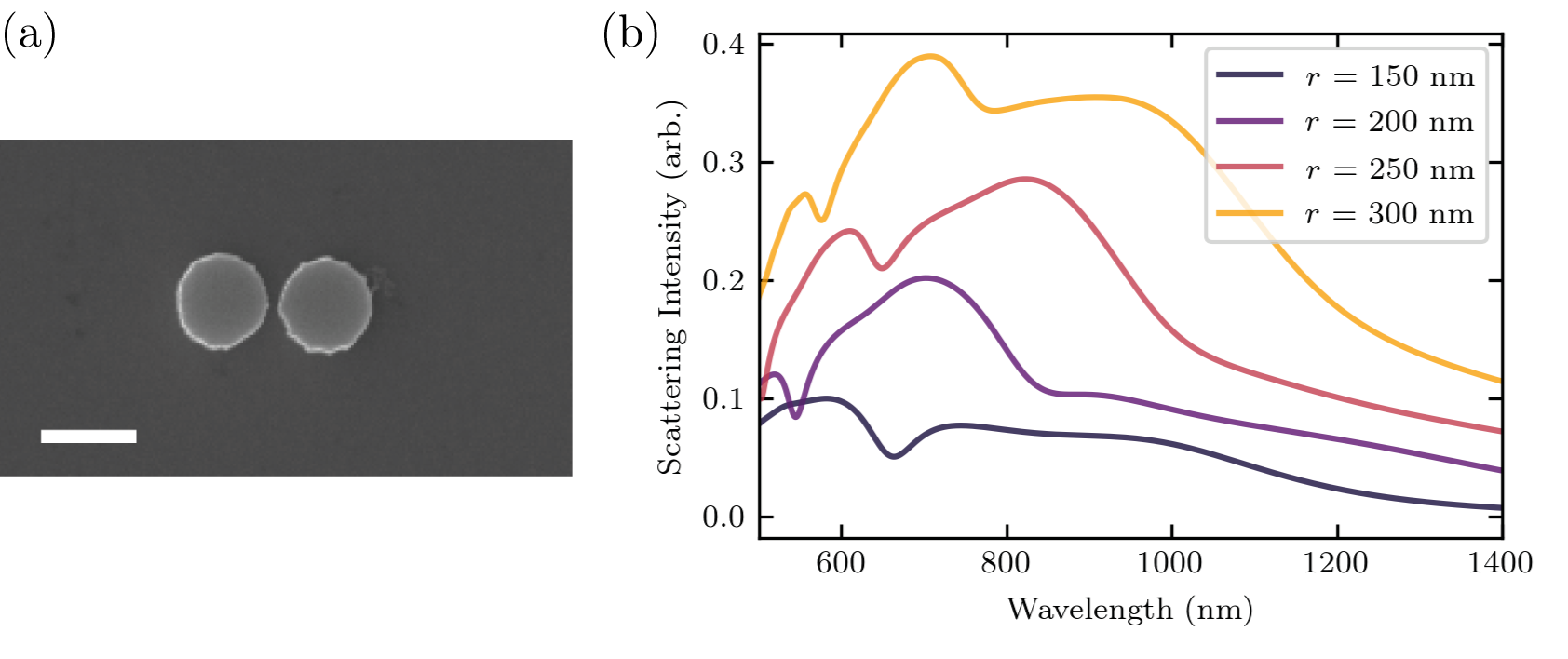}
	\caption{(a) Electron microscope image of a GaP dimer nano-antenna ($ r = 200  $ nm, $ h =200 $ nm). Scale bar = 200 nm. (b) Simulated scattering cross section for GaP dimer nano-antennas with varying radius $ r=150,200,250,300 $ nm, height $ h=200$ nm and gap width $ g=50 $ nm)}
	\label{fig:siscattering}
\end{figure}

\clearpage

\section*{Supplementary note III: Photoluminescence and polarization properties of WSe$_2$ single photon emitters on GaP nano-antennas}

\begin{figure}[b!]
	\centering
	\includegraphics[width=1\linewidth]{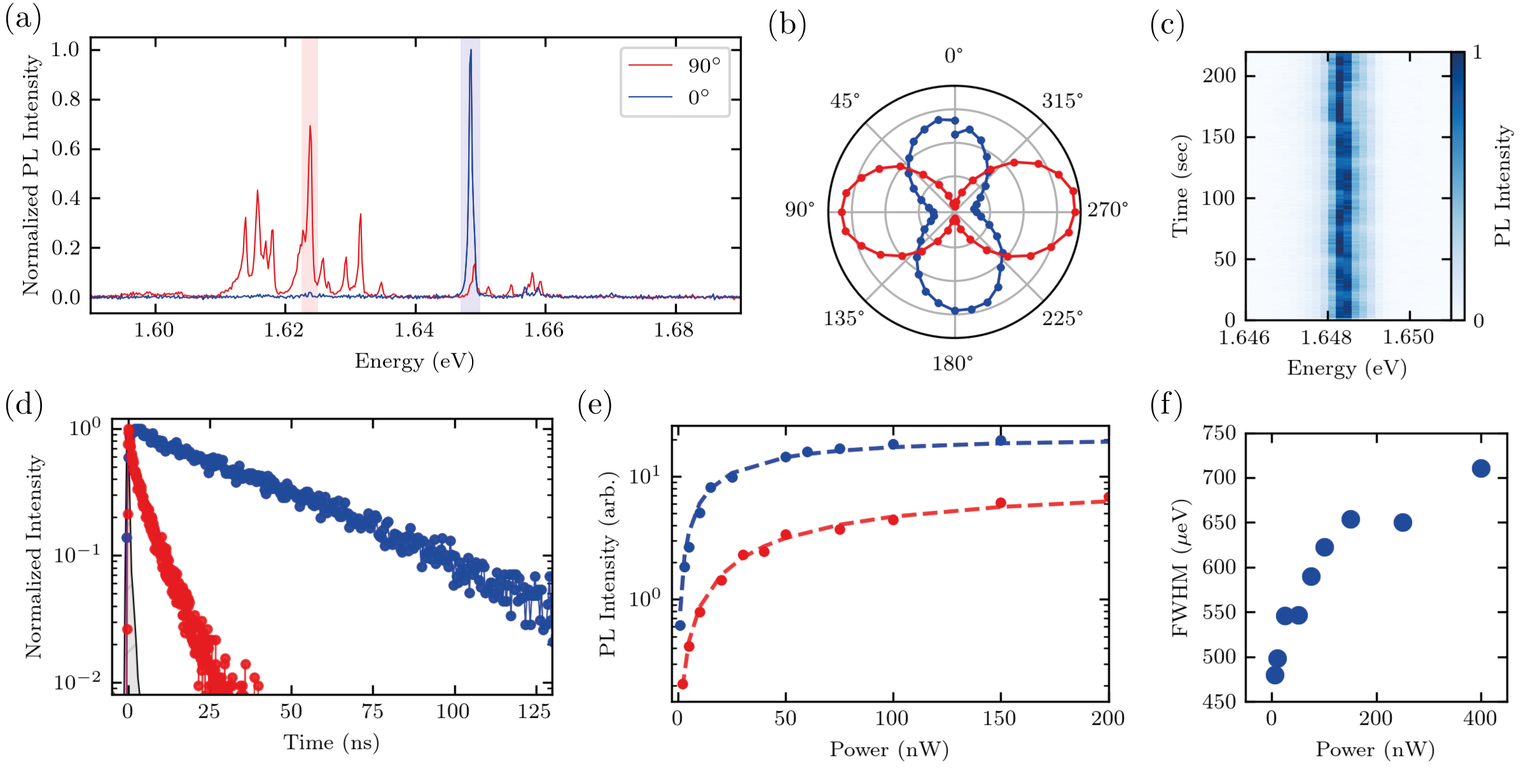}
	\caption{(a) Monolayer WSe$_2$ PL spectrum, collected on top of a single GaP nano-antenna at a temperature of $ T=4 $ K. The PL spectrum in blue is collected with a linear polarizer aligned to the excitation polarization axis, while the PL spectrum in red is recorded after rotating the linear polarizer in the detection path by 90 degrees. (b) Polar plot of the integrated PL intensity, for the peaks highlighted with the same colour in Supplementary Fig.\ref{fig:s1dotpl}a. (c) PL emission for the blue peak in Supplementary Fig.\ref{fig:s1dotpl}a showing stable emission over time. (d) Time resolved PL dynamics of the peaks highlighted with the same colour in Supplementary Fig.\ref{fig:s1dotpl}a. (e) Power dependent saturation of the PL intensity, under a pulsed excitation at 638 nm and 80 MHz repetition rate. A similar PL saturation threshold is found for both types of emitter.(f) Broadening of the linewidth (FWHM) under increasing power density for the co-polarized peak in Supplementary Fig.\ref{fig:s1dotpl}a.}
	\label{fig:s1dotpl}
\end{figure}

Supplementary Fig.\ref{fig:s1dotpl}a shows the PL spectrum for a WSe$ _2 $ monolayer on top of a GaP dimer nano-antenna ($ r=250 $ nm) excited with a linearly polarized laser at 638 nm, and collected under two orthogonal polarization directions (blue and red traces). As shown in Supplementary Fig.\ref{fig:s1dotpl}b, the single peak observed in co-polarized collection (highlighted in blue in Fig.S\ref{fig:s1dotpl}a) is accompanied by a series of peaks in cross-polarized detection at lower energy (highlighted in red in Fig.S\ref{fig:s1dotpl}a). The SPEs on GaP nano-antennas exhibit stable emission over time, as shown in Supplementary Fig.\ref{fig:s1dotpl}c, as expected from the improved spectral wandering in highly strained monolayers \cite{Palacios-Berraquero2016c}. 
As shown in Supplementary Fig.\ref{fig:s1dotpl}d, we observe ns PL lifetimes, with faster dynamics for the cross-polarized peaks (red trace $ \tau= $ 7 ns) compared to the co-polarized peak (blue trace $ \tau= $ 42 ns). 
The emitters exhibit a similar power saturation behaviour, as shown in Supplementary Fig.\ref{fig:s1dotpl}e. In Supplementary Fig.\ref{fig:s1dotpl}f we show the homogeneous broadening of the linewidth (FWHM) of the SPE highlighted in blue in Fig.S\ref{fig:s1dotpl}a, under increasing excitation power density, obtained by fitting the spectra with a Lorentzian peak function (see Fig.4 in the main text) . 
Due to the varying anisotropy of the confining potential, SPEs in WSe$ _2 $ are known to exhibit emission peaks both with and without a fine structure splitting (FSS) at zero magnetic field \cite{Kumar2016}. As we show in Supplementary Fig.\ref{fig:s1dotplradius}a-b, we observe the presence of cross polarized peaks with FSS on the order of 600-800 $ \mu $eV ( Supplementary Fig.\ref{fig:s1dotplradius}c) and, on the same nano-antenna, peaks that show a near unity degree of linear polarization with no underlying fine structure.
As shown in Supplementary Fig.\ref{fig:s1dotplradius}d-f, we observed a repeated pattern in the polarization from strain-induced SPEs on top of nano-antennas with different radii. The presence of a bright, high energy peak, usually exhibiting a FSS, is followed by a large number of emitters at lower energies, exhibiting the same polarization axis and no FSS. 
We ascribe this behaviour to different kinds of strain-induced emitters in the WSe$ _2 $ layer, related to a different confinement energy and size of the potential well created by nano-scale strain inhomogeneities.
\begin{figure}[h!]
	\centering
	\includegraphics[width=1\linewidth]{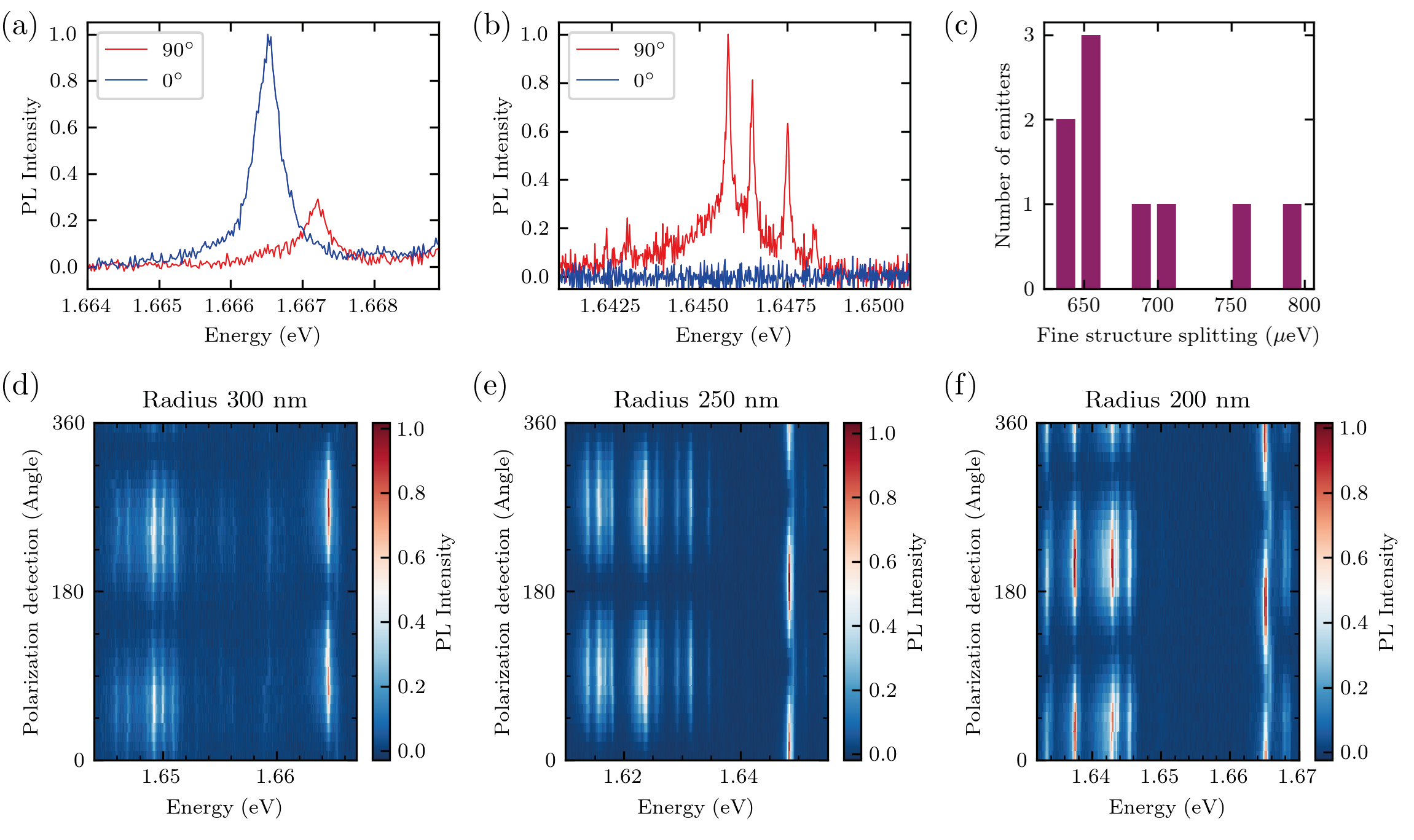}
	\caption{High resolution PL spectra at co- and cross-polarized detection centered at the blue (a) and red (b) highlighted peaks in Supplementary Fig.\ref{fig:s1dotpl}a. Supplementary Fig.\ref{fig:s1dotplradius}a displays a fine structure split cross-polarized doublet. (b) does not display any FSS and exhibits a near unity degree of linear polarization. (c) Histogram of fine structure splitting values observed for high energy peaks at the positions of different nano-antennas. (d-e) PL emission as a function of the detection polarization angle for localized SPEs on top of GaP dimer nano-antennas with different radii of 300 nm (d), 250 nm (e) and 200 nm (f).}
	\label{fig:s1dotplradius}
\end{figure}

\clearpage

\section*{Supplementary note IV: Fabrication and photoluminescence properties of WSe$ _2 $ single photon emitters on SiO$ _2 $ nano-pillars}

\begin{figure}[h!]
	\centering
	\includegraphics[width=1\linewidth]{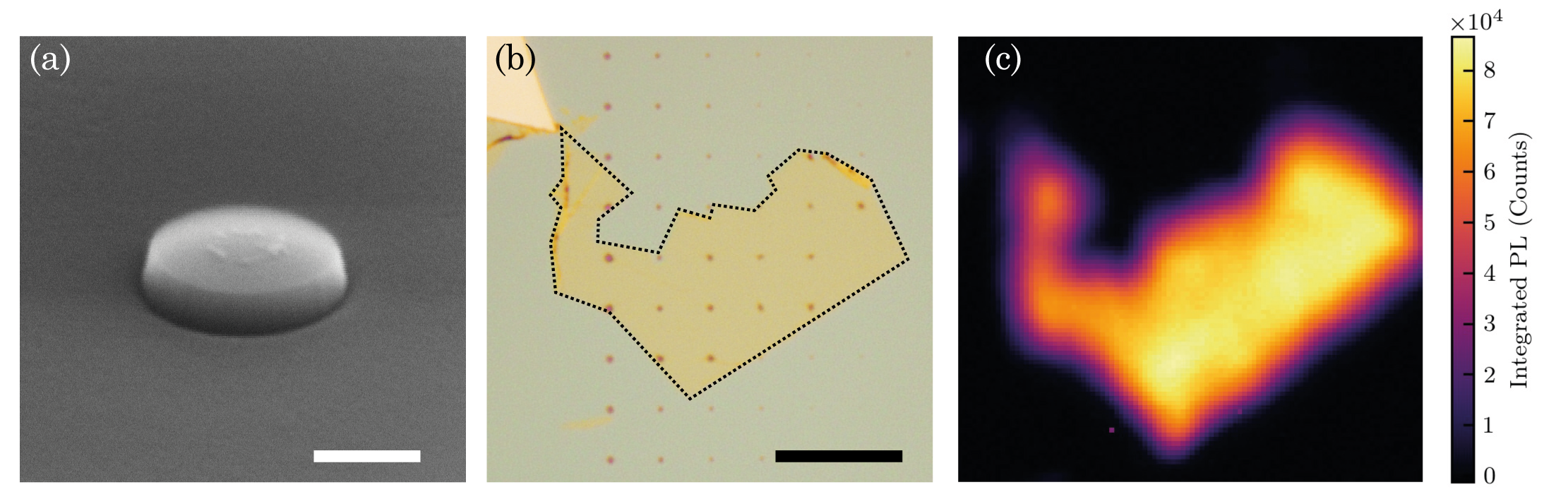}
	\caption{(a) Electron microscopy tilted image of a SiO$ _2 $ nano-pillar ($ h\approx 100 $ nm). Scale bar: 200 nm.  (b) Bright field optical microscopy image of a monolayer WSe$ _2 $ (outlined by the dashed line) transferred on top of the array of SiO$ _2$ nano-pillars. Scale bar: 10 $ \mu $m. (c) Room temperature PL intensity map of the transferred monolayer WSe$ _2 $. }
	\label{fig:s1sio2pillars}
\end{figure}

The SiO$_2$ nano-pillars are fabricated from a thermally grown 290 nm SiO$_2$ layer on a silicon wafer, with radii ranging from 50 to 250 nm, with an electron beam lithography and reactive ion etching system. Supplementary Fig.\ref{fig:s1sio2pillars}a shows an electron microscope image of a resulting SiO$ _2 $ nano-pillar with a radius of 200 nm and height of 100 nm.
The monolayer of WSe$_2$ is transferred onto the SiO$_2$ nano-pillars with an all-dry transfer technique. A bright field image of the transferred monolayer is shown in Supplementary Fig.\ref{fig:s1sio2pillars}b. Supplementary Fig.\ref{fig:s1sio2pillars}c shows the room temperature PL map of the transferred WSe$ _2 $ monolayer on top of the nano-pillar array.

\begin{figure}[h!]	
	\centering
	\includegraphics[width=1\linewidth]{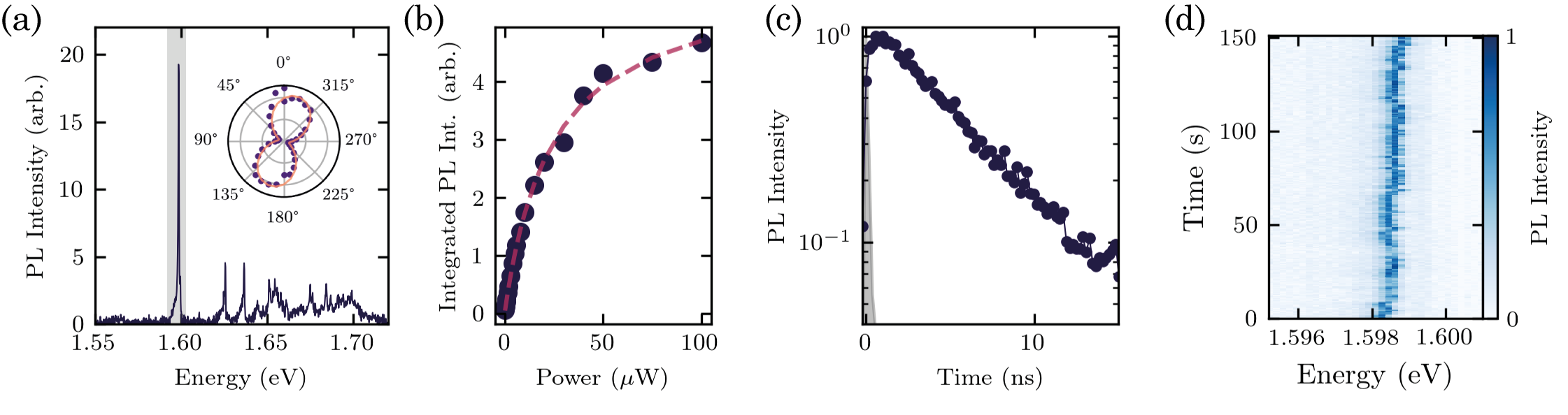}
	\caption{(a) PL spectrum of a monolayer WSe$ _2 $ SPE on top of a SiO$ _2 $ nano-pillar, collected at a temperature of $ T=4 $ K. Inset: polar plot of the linearly polarized emission for the SPE peak highlighted in grey. (b) Power saturation of the PL intensity under pulsed excitation at 638 nm and 80 MHz repetition rate. (c) Time resolved luminescence of the SPE localized on a SiO$ _2 $  nano-pillar (excitation pulse in grey). (d) Temporal stability of the PL emission from a SPE positioned on a SiO$ _2 $ nano-pillar ($ h=100 $ nm). }  
	\label{fig:s1sio2spe}
\end{figure}

Supplementary Fig.\ref{fig:s1sio2spe}a shows a representative cryogenic PL spectrum of a WSe$_2$ SPE positioned on top of a SiO$ _2 $ nano-pillar (highlighted in grey). The localized emitter exhibits linearly polarized emission (Inset Supplementary Fig.\ref{fig:s1sio2spe}a), as expected from 2D in-plane dipole emitters. The SPE further shows saturation of the PL intensity under increasing excitation power (Supplementary Fig.\ref{fig:s1sio2spe}b),  PL decay with $ \tau = 5.8 $ ns (Supplementary Fig.\ref{fig:s1sio2spe}c) and lower spectral stability compared to SPEs on GaP nano-antennas (Supplementary Fig.\ref{fig:s1sio2spe}d).

\clearpage

\section*{Supplementary Note V: Strain dependence of the single photon emission}
\begin{figure}[h!]
	\centering
	\includegraphics[width=1\linewidth]{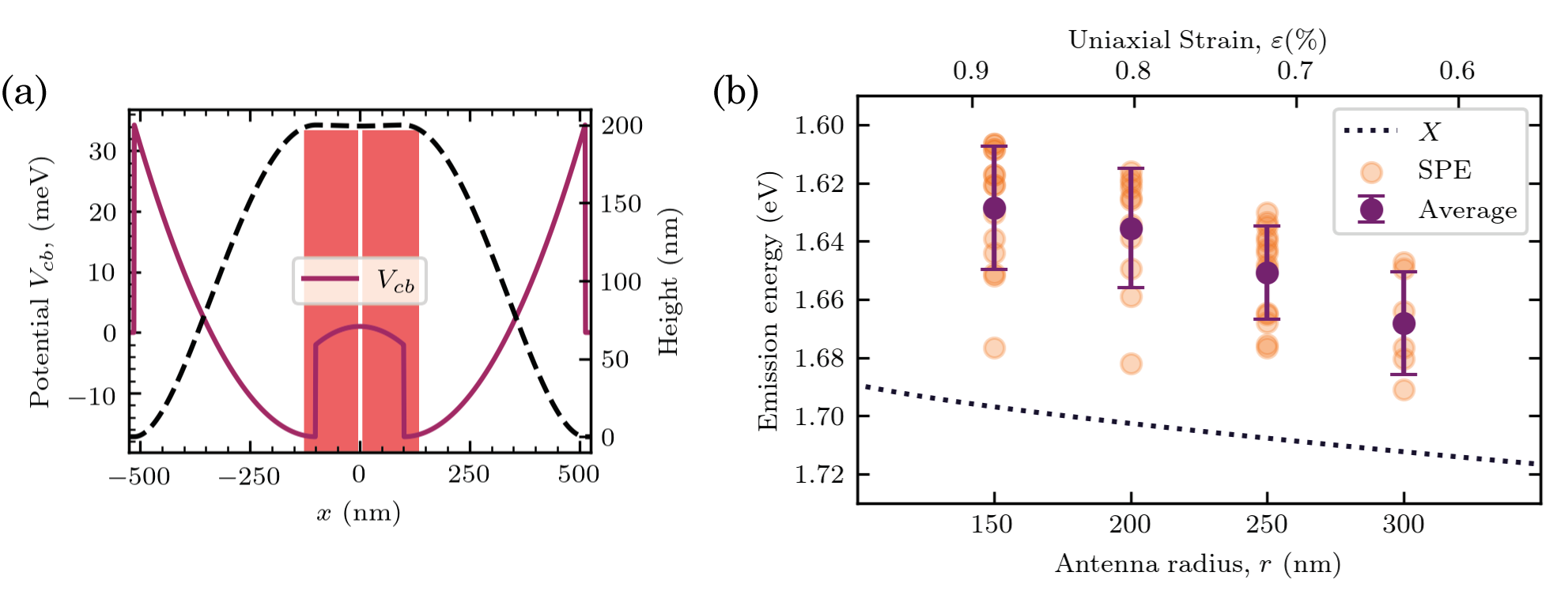}
	\caption{(a) Cross-section along the x-axis (as in Figure 1 of the main text) of the strain-induced conduction band potential ($ V_{cb} $) modulation (in purple) calculated for the K valley of a monolayer WSe$ _2 $ as a function of x. The black dashed line shows the heigth profile for the same WSe$ _2 $ monolayer on top of the dimer nanoantenna (in red) \cite{Sortino2020}. 
		(b) Emission energy of WSe$ _2$ SPEs deposited on GaP nano-antennas with different radii (yellow dots), compared with the energy red-shift of the free exciton in WSe$ _2$ ($X$, dashed line) as a function of the nano-antenna radius. In purple the average value, the error bars show the standard deviation. The dashed line is calculated from the unstrained value and obtained by interpolating the theoretical curve described in Ref.\cite{Sortino2020} with the experimental gauge of -49 meV/\% under tensile strain for the WSe$ _2 $ exciton at room temperature from Ref.\cite{Niehues2018a}. }
	\label{fig:scatterstrain}
\end{figure}

The nano-antenna geometry can be used to tailor the strain introduced in an atomically thin semiconductors as described in Ref.\cite{Sortino2020}. Supplementary Fig.\ref{fig:scatterstrain}a shows the section along the x-axis, defined as in Fig.1 of the main text, of the height profile (black dashed line) of a WSe$ _2 $ monolayer on top of a dimer nano-antenna (in red) and the relative change in the WSe$ _2 $ conduction band potential ($ V_{cb} $). The tensile strain is maximized at the edges of the nano-pillars (in red) and correspond to a lowering of $ V_{cb} $, and directly a reduction of the band gap energy. This strain profile forms a deformation potential well which can trap excitons \cite{Sortino2020}. Where the 2D layer touches the substrate, strain becomes compressive and the $ V_{cb} $ is increased.

In Supplementary Fig.\ref{fig:scatterstrain}b we show the position of different SPEs emission wavelengths (orange dots) and their average (purple), as a function of the nano-antenna radius. The increasing red-shift of the SPEs emission energy when on smaller radii nano-antennas is related to an increased tensile strain introduced in the 2D-WSe$ _2 $ membrane.
These results confirm the impact of strain on the emission properties of strain-induced SPEs in two-dimensional WSe$ _2 $.

\clearpage

\section*{Supplementary note VI: Collection and quantum efficiency of the single photon emission}

The underlying quantum efficiency ($ QE $) of a SPE under pulsed excitation can be estimated from the laser repetition rate and the number of detected photons \cite{Luo2018}. If for each laser pulse we detect a photon, the $ QE $ is 100\% and the rate of photons detected matches that of the excitation laser repetition rate.
We calibrated the collection efficiency by measuring the losses with a 725 nm laser. The values obtained from the calibration are listed in the table below. The values for the transmission of the linear polarizer, the spectrometer and the CCD efficiency are taken from the relative datasheet. We obtain a collection efficiency of the experimental setup of 0.56$ \% $.
\begin{table}[h!]
	\begin{tabular}{|l|l|lll}
		\cline{1-2}
		\textit{Component}                   & \textit{Transmission} &  &  &  \\ \cline{1-2}
		Optical components (Cryostat)                  & 50\%                  &  &  &  \\ \cline{1-2}
		Linear polarizer                     & 78\%                  &  &  &  \\ \cline{1-2}
		Single Mode fibre coupling           & 2\%                   &  &  &  \\ \cline{1-2}
		Spectrometer in-coupling             & 90\%                  &  &  &  \\ \cline{1-2}
		Spectrometer mirrors (x3)            & (97\%)$^3$=91\%                 &  &  &  \\ \cline{1-2}
		Grating Efficiency                   & 90\%                  &  &  &  \\ \cline{1-2}
		CCD Quantum Efficiency               & 98\%                  &  &  &  \\ \cline{1-2}
		\textbf{TOTAL COLLECTION EFFICIENCY} & \textbf{0.56\%}      &  &  &  \\ \cline{1-2}
	\end{tabular}
\end{table}

\begin{figure}[h!]
	\centering
	\includegraphics[width=0.6\linewidth]{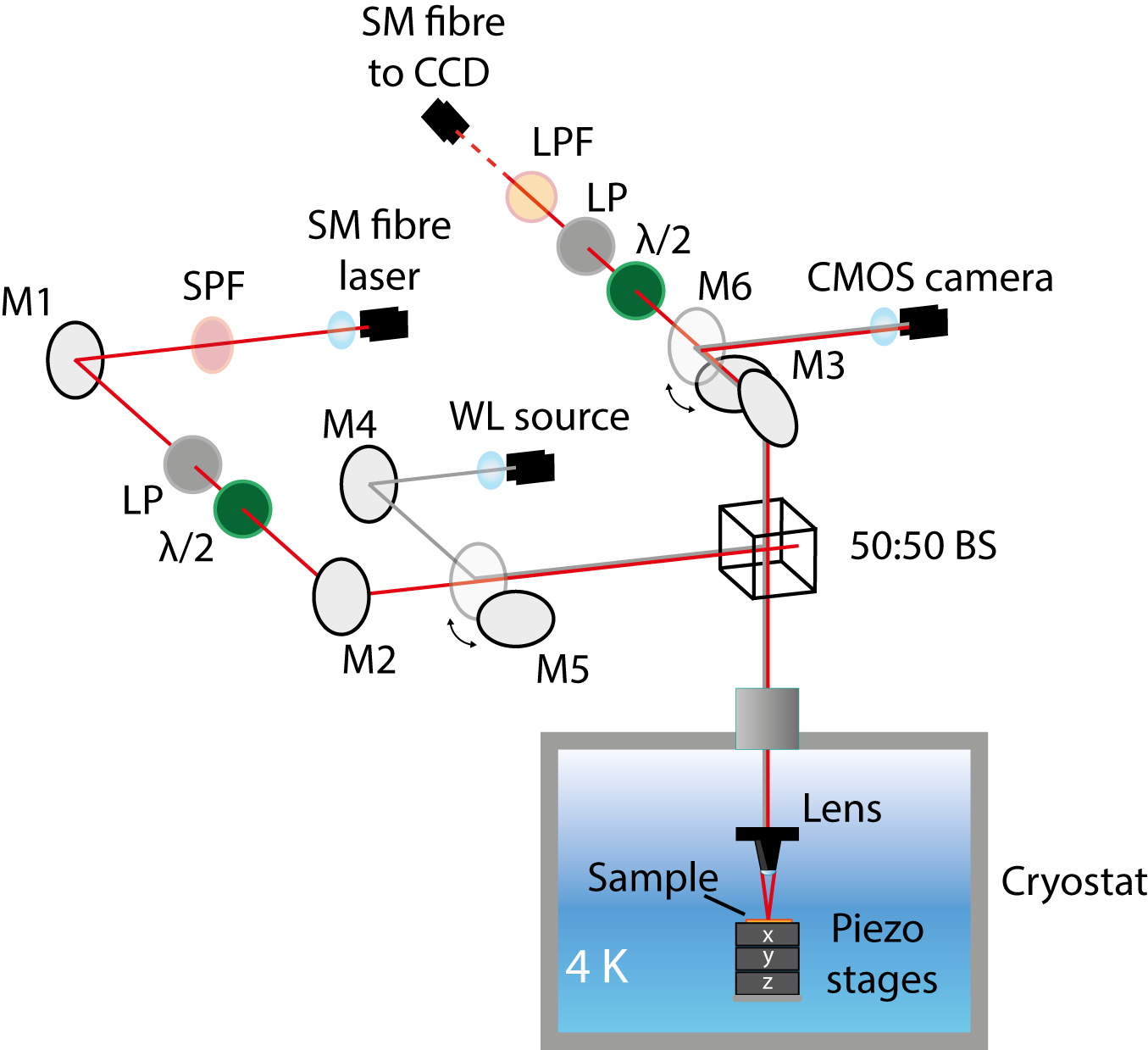}
	\caption{Schematics of the experimental setup used to excite and collect the light from the SPEs placed in a bath cryostat at T = 4 K. SM: single mode fibre, SPF: short pass filter, M: mirror, LP: linear polarizer, $ \lambda/2 $: half-wave plate, WL source: white light source, BS: beam splitter, LPF: long pass filter.}
	\label{fig:exp}
\end{figure}

From the FDTD simulations described in Supplementary Note I, we have estimated the cavity collection efficiency as the fraction of the total power radiated inside an objective with NA=0.64, the same used in our experiments. The estimated internal quantum efficiency ($ QE $) is calculated as:
\begin{equation}
	QE = \dfrac{I\cdot\eta^{-1}_{exp}\cdot\eta_{coll}^{-1}}{R}
\end{equation}
\noindent where $ I $ is the collected intensity from the SPE, $ \eta_{exp} $ is the collection efficiency of the experimental setup, $ \eta_{coll} $ is the collection efficiency from numerical simulations (given in Supplementary Note I), and $ R $ is the laser repetition rate.
In Supplementary Fig.\ref{fig:qehist} we show the estimated internal quantum efficiency for WSe$ _2 $ SPEs positioned on both GaP dimer nano-antennas and SiO$ _2 $ nano-pillars, corrected for the total collection efficiency and for the corresponding laser repetition rate. For SiO$ _2 $, we obtain an average quantum efficiency of $ 4\% $ while for SPEs on GaP an average of $ 21\% $ with some SPEs reaching values as high as $ 86\% $, corresponding to an overall single photon emission rate of 69 MHz and, normalized to the collection efficiency, of a single photon rate of 5.5 MHz at the first lens.
\begin{figure}[h!]
	\centering
	\includegraphics[width=0.65\linewidth]{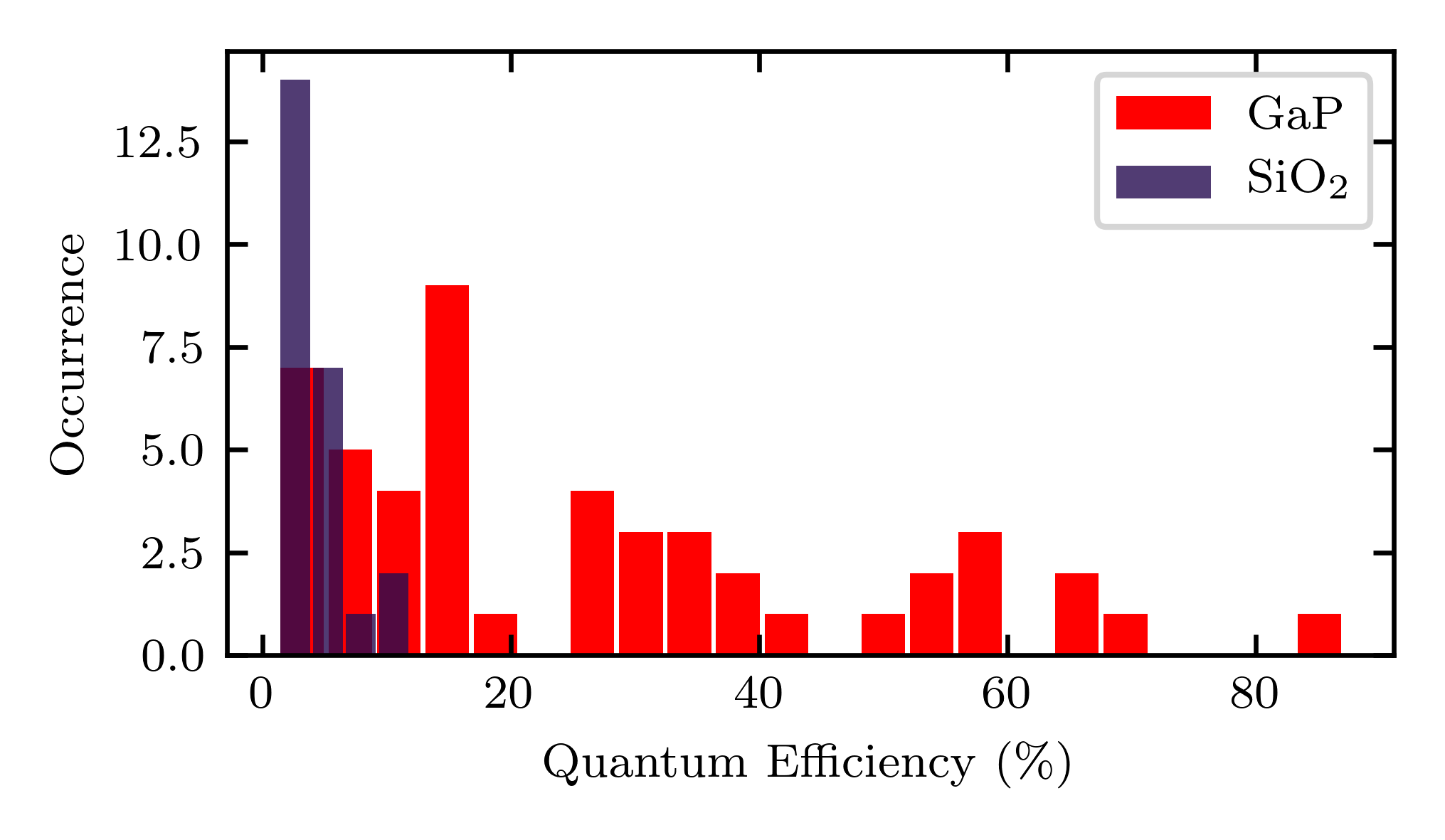}
	\caption{Estimated internal quantum efficiency of the single photon emission in WSe$ _2 $ SPEs on GaP nano-antennas (red) and on SiO$ _2 $ nano-pillars (blue).}
	\label{fig:qehist}
\end{figure}

\newpage
\pagebreak

\section*{Supplementary note VII: Photoluminescence dynamics of strain-induced WSe$ _2 $ single photon emitters}
To obtain an analytical solution of the SPEs PL dynamics with a three level system, as discussed and depicted in Fig.3d in the main text, we use a reduced version of the model which can be solved analytically, given by the following equations: 
\begin{align}
	\dfrac{dn_1}{dt} =-\dfrac{n_1}{\tau_1}+\dfrac{n_X}{\tau_2}  &&
	\dfrac{dn_X}{dt} = -\dfrac{n_X}{\tau_2}
\end{align}
\noindent
where $ n_1 $ and $ n_X $ are the populations of the SPE and dark excitons, respectively. Here, $ \tau_1 = (\Gamma_{r}+\Gamma_{nr})^{-1}$ is the recombination process of the SPE state giving rise to the luminescence, considering both radiative and non-radiative processes, while $ \tau_2 = (\Gamma_{trap}+\Gamma_{nr}^{X})^{-1}$, composed of both trapping rate of a single exciton into the strain-induced potential and the non-radiative decay of the exciton population . We exclude from this model the quadratic Auger term and the saturation of the dot discussed in the main text. The analytical solution to the above equation system is given by:
\begin{align}
	n_1(t) = e^{-\frac{t}{\tau_1}} [n_1(0)+ A(e^{\frac{t}{\tau_1}-\frac{t}{\tau_2}}-1)]
\end{align}
\noindent
where $A = n_X(0)\frac{\tau_1}{\tau_2-\tau_1} $, and $ n_X(0) $ and $ n_1(0) $ are the initial conditions for each relative rate equation. The above equation is used to fit the experimental data and obtain the values of the rise time ($\tau_{rise} = \tau_2$) and decay time ($ \tau_{decay}=\tau_1 $) shown in the main text.

\begin{figure}[h!]
	\centering
	\includegraphics[width=0.9\linewidth]{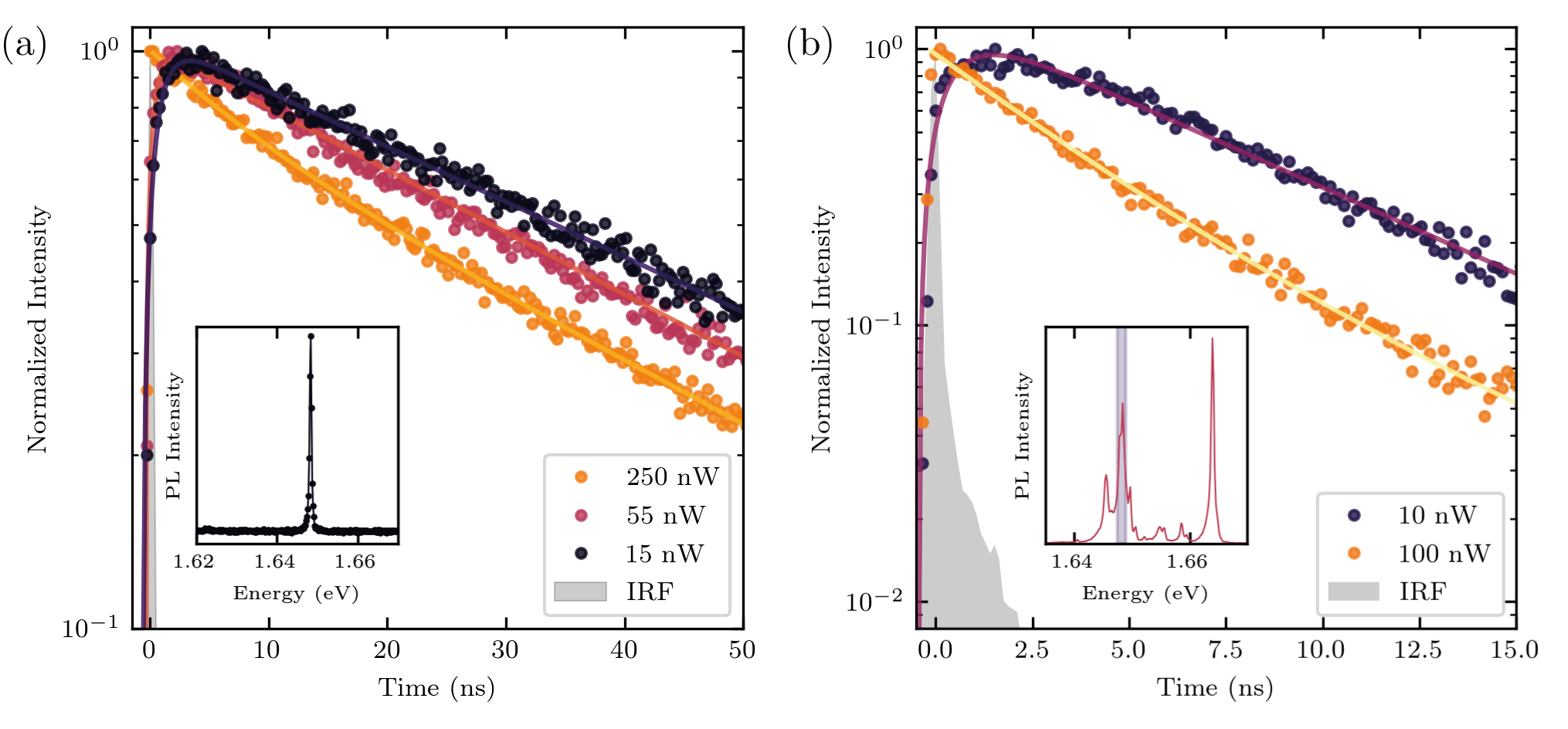}
	\caption{(a) Power dependence of the PL lifetimes for the SPE shown in Figure 1 of the main text (see Inset). (b) Power dependence of the PL lifetimes for the side peaks of the SPE shown in Figure 3 of the main text (highlighted in grey in the Inset). In grey, the instrument response function (IRF).}
	\label{fig:s13level}
\end{figure}

\noindent
Supplementary Fig.\ref{fig:s13level} shows additional PL decay from different SPEs, the PL spectra of which is shown in the figure inset, and fitted with the analytical solution of the model describe above. Under increasing power, the rise time reduces below the timing resolution of our experimental setup, and the PL decay becomes a single exponential profile.

\end{document}